\documentclass[aps,reprint, prx, superscriptaddress, showpacs,nobibnotes,nofootinbib]{revtex4-1}
\usepackage{graphicx}
\usepackage{bm,amsmath}
\usepackage{systeme}
\usepackage{hyperref}
\usepackage{mathrsfs}
\usepackage{dcolumn}
\usepackage{natbib}
\usepackage{braket}
\usepackage[dvipsnames]{xcolor}
\usepackage{ulem}
\usepackage[version=4]{mhchem}
\usepackage[title]{appendix}
\usepackage{blkarray}

\begin{document}
\title{Slow spin-lattice relaxation dynamics in YbVO$_4$ revealed by extended thermal impedance spectroscopy from AC susceptibility and AC magnetocaloric measurements}

\author{Yuntian Li}
\affiliation{Department of Applied Physics, Stanford University, Stanford, California 94305, USA}
\affiliation{Geballe Laboratory for Advanced Materials, Stanford University, Stanford, California 94305, USA}
\author{Jiayi Hu}
\affiliation{Department of Applied Physics, Stanford University, Stanford, California 94305, USA}
\affiliation{Geballe Laboratory for Advanced Materials, Stanford University, Stanford, California 94305, USA}
\author{Dominic Petruzzi}
\affiliation{Department of Physics, Stanford University, Stanford, California 94305, USA}
\affiliation{Geballe Laboratory for Advanced Materials, Stanford University, Stanford, California 94305, USA}
\author{Linda Ye}
\affiliation{Department of Applied Physics, Stanford University, Stanford, California 94305, USA}
\affiliation{Geballe Laboratory for Advanced Materials, Stanford University, Stanford, California 94305, USA}
\author{Mark P. Zic}
\affiliation{Geballe Laboratory for Advanced Materials, Stanford University, Stanford, California 94305, USA}
\affiliation{Department of Physics, Stanford University, Stanford, California 94305, USA}

\author{Arkady Shekhter}
\affiliation{Los Alamos National Laboratory, Los Alamos, New Mexico 87545, USA}
 
\author{Ian R. Fisher}
\affiliation{Department of Applied Physics, Stanford University, Stanford, California 94305, USA}
\affiliation{Geballe Laboratory for Advanced Materials, Stanford University, Stanford, California 94305, USA}

\date{\today}

\begin{abstract}

 Alternating (AC) magnetic fields can induce not only an alternating magnetization in materials, but also an alternating temperature via the magnetocaloric effect. The latter effect is typically neglected when performing AC susceptibility measurements, but consideration of both effects on an equal footing is necessary in order to reliably distinguish between internal and external causes of magnetic response and accurately extract quantitative information about relaxation processes. In order to address this, we have developed a method to measure the AC magnetocaloric effect that is compatible with AC susceptibility measurements, and also a framework to analyze these data in combination. We demonstrate the efficacy of this approach using \ce{YbVO4}, a material for which strong single-ion anisotropy leads to slow spin-lattice relaxation at low temperatures via a phonon bottleneck effect. We report AC magnetic susceptibility and AC magnetocaloric effect measurements for this material as a function of field and frequency at a temperature of 3 K. We analyze the data using a discretized thermal model, and extract the field-dependence of the intrinsic spin-lattice relaxation rate.  This demonstration experiment illustrates a general approach to quantitatively address multiple measured quantities in driven systems using a unified thermal circuit analysis. The thermal analysis methods presented in this report can be extended to study other magnetic, dielectric, and elastic materials exhibiting a complex response to an external driving field in the presence of internal and external relaxation, particularly when an energy dissipation process is within an accessible frequency regime.

\end{abstract}

\maketitle

\section{Introduction}

\begin{figure}[ht!]
	\includegraphics[width = \columnwidth]
    {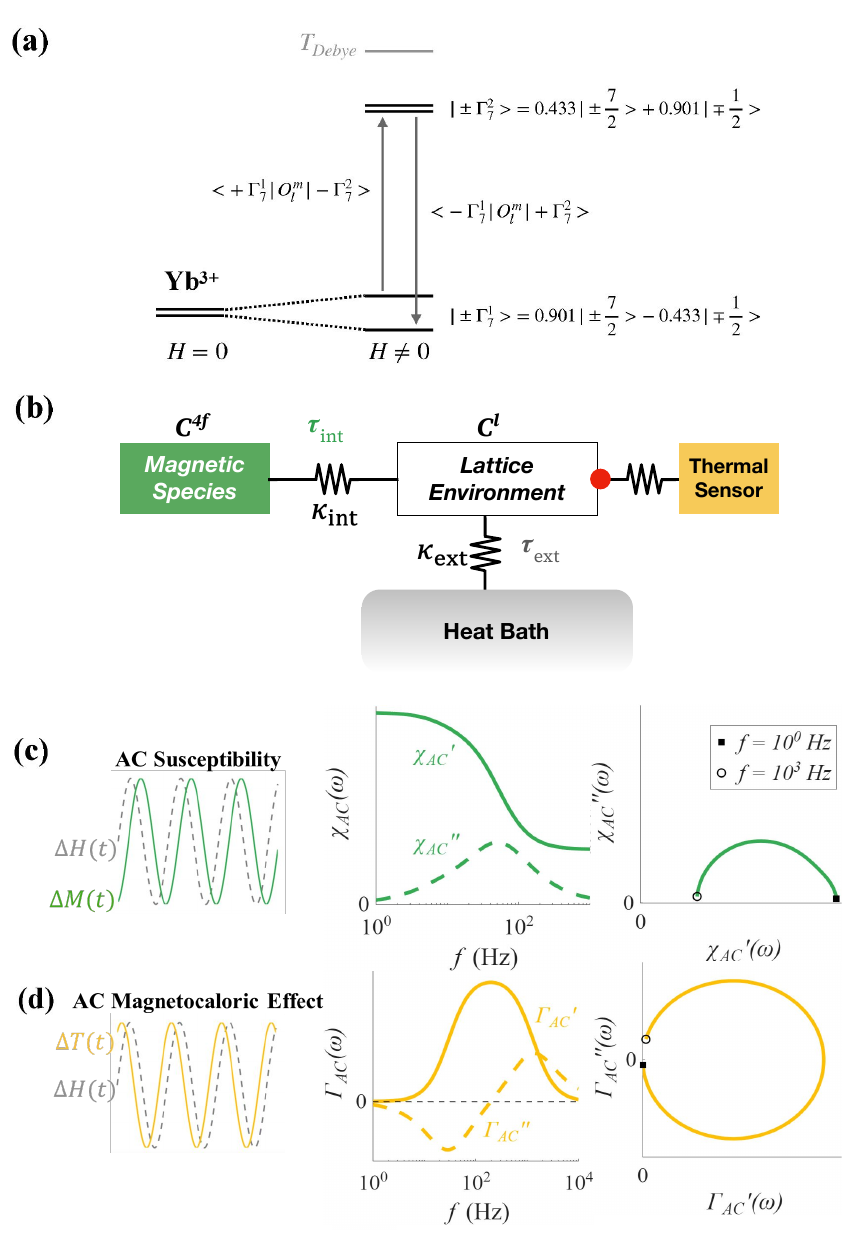}
	\caption{\label{fig-1} (a) Schematic diagram illustrating the CEF levels and indirect transition corresponding to \ce{Yb^{3+}} ion embedded in the \ce{YbVO4} crystal field environment. The lower two levels represent the ground state Kramers doublet, which are split when an external magnetic field is oriented along the c-axis. The higher energy states correspond to the excited states of the crystal field levels. We consider a scenario where the temperature is much less than the Debye temperature. (b) Schematic diagram illustrating the components of the discretized thermal circuit model considered in this report. The coefficients and time constants are defined in the main text. (c,d) Schematic diagrams illustrating solutions of the equivalent thermal circuit, and using parameters characteristic of YbVO$_4$ at 3K. In the left panels, the normalized response is plotted to illustrate the phase difference between the time-dependent solution and the signals. In the middle panels, solid lines represent the real components ($\chi_{AC}'$ and $\Gamma_{AC}'$), and dashed lines represent the imaginary components ($\chi_{AC}''$ and $\Gamma_{AC}''$). In the right panels, the imaginary components of each solution are plotted against their respective real parts in a Cole-Cole plot. This representation maintains a fixed aspect ratio of 1:1 for the $x$ and $y$ axes.}
\end{figure}

AC magnetic susceptibility, $\chi_{AC}$, describes the magnetic response of a material to an oscillating magnetic field \cite{Topping_2019}. The zero frequency limit defines the isothermal susceptibility, $\chi_T$, for which all the internal subsystems (spins, phonons, etc) are in equilibrium. For non-zero frequencies, $\omega$, the spin and lattice subsystems may not have enough time to equilibrate and the response can acquire a frequency dependence, $\chi_{AC}(\omega)$. A classic example is the phonon-bottleneck effect that is observed at low temperatures in many magnetic materials that exhibit magnetic anisotropy \cite{Orbach1961}. Oscillating magnetic fields also necessarily result in a thermal response due to the magnetocaloric effect (MCE). Thermal relaxation to the bath, as well as internal relaxation processes, are typically neglected in magnetic measurements because thermal equilibration is assumed to be fast \cite{Quilliam2008, Quilliam2011, Coca2014}. However, in systems with slow spin dynamics relative to the driving frequency, this assumption is unwarranted. To account for magnetization-induced heat dynamics, a full treatment of the combined magnetic and thermal response is required. Here, we provide such a framework to analyze the AC magnetic susceptibility in concert with the AC magnetocaloric response of materials. Our approach builds upon earlier work \cite{Khansili2023} to develop an extended thermal impedance spectroscopy that fully describes the effective magnetic and thermal response functions of the material coupled to a thermal bath. We also describe a practical experimental method to measure the AC MCE that is compatible with standard AC magnetic susceptibility experiments. 

We specifically choose the magnetic insulator \ce{YbVO4} to demonstrate these effects. In this material, the Crystal Electric Field (CEF) splits the Hund's rule ground state multiplet of the Yb$^{3+}$ ion over a wide energy range, resulting in a Kramers doublet ground state with the first excited state an equivalent of 91.7 K higher in energy (see Fig. \ref{fig-1}(a)) \cite{Bowden1998}. The effective interaction between nearby ions is primarily via dipolar interactions, resulting in a relatively low N\'{e}el temperature of just 93 mK \cite{Radhakrishna1981}. As we will show, the combination of strong magnetic anisotropy (large CEF splitting), weak coupling between the moments, and weak coupling to phonons (due to the Kramers nature of the doublet groundstate), make the dynamics of the \ce{Yb} moments slow relative to typical driving frequencies at low temperatures, and hence render this an especially useful material to demonstrate our approach.

The relevant internal dynamics affecting $\chi_{AC}(\omega)$ in \ce{YbVO4} are governed by the spin-lattice relaxation rate. An applied magnetic field lifts the degeneracy of the Kramers ground state doublet of the \ce{Yb} ions. If this is done fast enough relative to the characteristic spin-lattice relaxation time, occupation numbers of the distinct CEF states cannot be changed rapidly enough, and the combined spin and lattice system is no longer in thermal equilibrium. Spin-lattice relaxation occurs by transitions between the distinct CEF eigenstates and is either direct (involving a single phonon) or indirect (the transition involves  intermediate states and two or more phonons). For Yb ions in \ce{YbVO4}, direct transitions are not allowed by symmetry (see Appendix. \ref{app:CEF}), and the ion can only relax by indirect processes (Fig. \ref{fig-1}(a)). Zeeman splitting of the CEF eigenstates necessarily leads to a strong field dependence of the associated relaxation rates, making this an especially useful material system to identify internal relaxation effects since these can be readily tuned. 

Spin-lattice relaxation time scales can be surprisingly long in rare-earth compounds that exhibit large magnetic anisotropies \cite{Orbach1961}. In \ce{YbVO4}, at a temperature much lower than the Debye temperature of the lattice, very few phonons are available to participate in the dynamics, resulting in a low transition probability -- a manifestation of the phonon bottleneck effect. The indirect nature of the transition effectively decouples the material into separate baths of spins and phonons (Fig. \ref{fig-1}(b)) for quite modest frequencies, with a long internal time constant $\tau_{int}$ that describes the non-instantaneous relaxation between these two baths. The magnitude of $\tau_{int}$ can be related to an effective internal heat conduction rate $\kappa_{int}$ (where $\kappa_{int}$ is different from the lattice thermal conductivity and should not be confused with it) and the heat capacities associated with the spin and lattice degrees of freedom, $C_{4f}$ and $C_l$ respectively:  $\tau_{int} = \frac{C^lC^{4f}}{\kappa_{int}(C^l+C^{4f})}$ \footnote{This can be readily derived by considering the thermal relaxation between two objects with different heat capacity, which results in an exponential relaxation with the characteristic relaxation time that is given.}. As described below, we find characteristic frequencies of order 10 to 100 Hz at a temperature of 3 K, with the value depending on the value of the applied magnetic field, following an exponential functional form.  

The spin subsystem can be assigned a well-defined temperature on these timescales because internal equilibration processes are sufficiently fast to establish thermal equilibrium within the spin system on a shorter timescale. In analogy with nuclear spins \cite{Abragambook}, dipolar interactions between Yb spins, although weak, can be effective because the spin splittings are nearly resonant with each other. Moreover, the external perturbations -- the applied magnetic field and the spin–lattice temperature difference -- are spatially uniform across the sample, which leads to a uniform spin temperature in our experimental geometry even in the absence of additional internal equilibration processes within the spin subsystem.

In addition to internal relaxation processes, external relaxation can also affect $\chi_{AC}(\omega)$. Here, the issue is related to the flow of energy between the material and the environment, which is typically idealized as a heat bath (Fig. \ref{fig-1}(b)). Whereas the internal relaxation is intrinsic to any given material, external relaxation, characterized by a timescale $\tau_{ext}=\frac{C^{l}}{\kappa_{ext}}$ (where $\kappa_{ext}$ characterizes thermal conduction from the lattice to the bath) that can vary depending on experimental details. External relaxation must be accounted for if the internal relaxation time is to be quantitatively determined by any experiment. Furthermore, external relaxation plays a significant role in the effective MCE response of materials.  

MCE refers to the temperature change in a material that responds to variations in an external magnetic field \cite{Warburg1881, Kohama2010}. It has been extensively studied and applied in various contexts, such as adiabatic demagnetization refrigeration techniques \cite{Bruck_2005} and in characterizing phase transitions \cite{Law2018,Pereira2024}. For rare earth containing materials in the paramagnetic state, large magnetic and MCE responses can be anticipated based on the associated large entropy of the spin system \cite{Kaze2006, Palacios2018}, and this is also true of \ce{YbVO4}. The dominant method that is typically used for MCE measurements is in the time domain, in response to a swept magnetic field (for a review, see, for example, \cite{Kohama2010}). While several previous studies attempted to measure the MCE in periodically changing fields (i.e. an AC, or dynamic, MCE) \cite{FISCHER199179, Tokiwa2011}, quantitative analysis of the response function has been limited \cite{ALIEV2016601, ALIEV2022169300}, and to date a generally-applicable formal analysis has not been available. In the present work we develop an appropriate thermal model that accounts for internal and external relaxation, and demonstrate a method to measure the AC MCE that is compatible with standard AC susceptibility tools.

The model that we develop to account for the coupled magnetic and thermal response functions is based on discrete thermal elements (magnetic species, lattice, thermal sensor, thermal bath) and neglects any/all thermal inhomogeneity that might be present in a real system. Hence, for this model to be applicable, experiments should be designed in such a way that the heat flow is described by a well-defined circuit where each element is in internal thermal equilibrium. In the present case, the characteristic timescale for thermal transport across the width of the sample is estimated to be tens of microseconds in the regime of interest  \footnote{The thermal diffusion time associated with  heat equilibrating across a sample with thickness $L$ is given by $\tau_D = (C^{vol}_P/\kappa)L^2$, where $C^{vol}_P$ is the volume heat capacity and $\kappa$ is the thermal conductivity. Using literature values for these quantities for the closely related material TmVO$_4$ at 3 K and 0 T of $C_P\simeq 0.4 \text{ J} \text{ mol}^{-1}\text{ K}^{-1}$ \cite{tmvo42022} (from which $C^{vol}_P \simeq 8.3\times10^3 \text{J}\text{ m}^{-3} \text{ K}^{-1}$) and $\kappa \simeq 0.35 \text{ W}\text{ K}^{-1}\text{ m}^{-1}$ \cite{Vallipuram2024} we obtain an estimate $\tau_D \simeq 6 \times 10^{-5}\text{ s}$ for a representative sample thickness $L \simeq 50~\mu$m. The estimate for $\tau_D$ will change in the presence of a magnetic field, which affects the heat capacity, but still remains fast relative to the characteristic relaxation times probed in our experiments for this material. Thus the assumption of discrete thermal elements is justified.}. Since we probe the material in a frequency range that is considerably slower than this, we are safe in our assumption of there being negligible thermal gradients across the sample, but this needs to be verified for any given material to be studied, and clearly delineates the regime in which this analysis is meaningful. 

The model that we develop also neglects glassy effects associated with non-ergodic systems. Such systems have more complex internal relaxation, beyond the scope of the present work. Nevertheless, we note that AC MCE measurements can still be anticipated to be a useful complement to AC susceptibility measurements even in such cases. 

The major finding of this report is summarized in Fig. \ref{fig-1}(c,d), where the two specific response functions $\chi_{AC}(\omega)$ (describing the magnetic response) and $\Gamma_{AC}(\omega)$ (describing the caloric response) can be derived by analytically solving a thermal circuit model, and can both be measured experimentally. We will use this formalism to extract the intrinsic spin-lattice relaxation time in YbVO$_4$ as a function of field at a temperature of 3K.

\section{Experimental methods}

Single crystals of \ce{YbVO4} and \ce{GdVO4} were synthesized via slow cooling in a flux of \ce{Pb2V2O7} using a mixture of high-purity rare-earth oxides precursors, \ce{Yb2O3} ($99.99\%$ purity from Alfa Aesar, CAS Number: 1314-37-0) and \ce{Gd2O3} ($99.99\%$ purity from Alfa Aesar, CAS Number: 12064-62-9). More details related to the flux-growth synthesis method can be found in Refs. \cite{feigelson1968flux, smith1974flux, oka2006crystal}. 

AC susceptibility measurements were performed in a Quantum Design Magnetic Property Measurement System (model MPMS-3) using the AC susceptometer measurement option, following standard protocols as described elsewhere \cite{Topping_2019}. \ce{YbVO4} grows as long, thin crystals, with the c-axis directed along the long axis of the crystal. Measurements were performed with the magnetic field oriented along this direction. Demagnetization effects were minimal given the high aspect ratio of the crystals used for the experiments. 

AC MCE was performed using a customized probe within the same MPMS-3, ensuring a direct comparison of the two techniques can be made. The MCE measurement device consists of a polished sample with a thickness between 30-100 $\mu m$, a thermal sensor attached to the top surface of the sample, and a quartz sample holder that is thermally anchored to the bottom surface of the sample. We employ a resistive sensor, in this instance choosing RuO$_x$ thin film sensors for their sensitivity in the regime of interest. The RuO$_x$ sensor is connected to a Wheatstone resistance bridge comprising three other identical RuO$_x$ resistors to improve sensitivity, with the other elements of the bridge thermally anchored to the heat bath (located on the part of the probe far apart from the measured material). Since the AC MCE results in an AC thermal signature, the signal averaging advantages of a lock-in technique can be used to measure the temperature oscillations. Moreover, since we use a resistive thermal sensor, we can employ a demodulation technique, effectively measuring the response at the sidebands formed from the combined effect of the AC current used to measure the resistance of the RuO$_x$ sensor (the carrier frequency) and the AC change in temperature associated with the AC magnetic field (the signal). A practical consideration is that the thermal sensor needs to have negligible magnetoresistance, at least for field variations corresponding to the amplitude of AC field; this is not an overly stringent requirement for AC amplitudes of a few Oe, and can be easily checked. Further details of the experiment, especially related to signal demodulation, are described in Appendix \ref{app:Thermometer}.

\section{Heat flow model} 
\label{sec: heat exchange model}  

In order to extend the analysis of MCE into the frequency domain, we introduce and develop a heat exchange model based on a discretized thermal analog circuit. This approach builds on similar recent work that utilized an AC thermal impedance analysis to access spin-lattice relaxation dynamics \cite{Khansili2023}, but here extended to cover complex caloric responses. The physical experiment necessarily introduces several characteristic time scales, associated with heat exchange between the sample and the bath, as well as heat exchange between the spin and phonon subsystems described above. To account for these, we first describe the complete thermodynamic model of the experimental setup. 

Fig. \ref{fig-1}(b) shows the schematic heat flow diagram. We apply a periodic magnetic field $H(t)$ and measure the magnetization of the \ce{Yb} 4f spin subsystem $M(t)$ as well as the temperature of the lattice $T^{lat}(t)$. As we apply a magnetic field, the magnetization and temperature of the 4f subsystem changes, bringing it out of thermal equilibrium with the lattice. This leads to heat flow between the 4f subsystem and the lattice, as well as between the lattice and the environment (idealized as a heat bath). For simplicity, we initially assume an idealized thermal sensor, which has zero heat capacity and is able to instantaneously measure the temperature of the crystal lattice. In Appendix \ref{app:Thermometer} we expand the thermal model to include a realistic thermal sensor that is also able to exchange heat with the sample.

The thermodynamics of the 4f spins is generated by their free energy (using $T$ and $H$ as variables), 
\begin{equation}\label{4f-free-energy}
\begin{aligned}
    dF^{4f} = -S^{4f}dT^{4f} - M^{4f}dH^{4f}, 
\end{aligned}
\end{equation}
where $S^{4f}$ and $M^{4f}$ are the 4f subsystem's entropy and magnetization respectively. Considering the partial derivatives of $S^{4f}$ and $M^{4f}$ with respect to temperature and field   \footnote{The cross relaxation matrix is obtained by considering the partial derivatives of $S^{4f}$ and $M^{4f}$, noting the standard Maxwell relation $\gamma^{4f}= \frac {\partial S^{4f}}{\partial H}=\frac{\partial M^{4f}}{\partial T}$. Thus $dS^{4f} = \frac{\partial S^{4f}}{\partial T}dT + \frac{\partial S^{4f}}{\partial H}dH = \frac{C^{4f}}{T_0}dT + \gamma^{4f}dH$. This yields the first half of equation \ref{4f_cross_relaxation}. The second half follows similarly by considering $dM^{4f}$.}, it is possible to write a matrix of the thermodynamic coefficients that describes the changes in the entropy and magnetization associated with the 4f electrons as a consequence of changes in temperature ($dT$) and field ($dH$) \cite{Landau}:

\begin{equation}\label{4f_cross_relaxation}
\begin{aligned}
   \left[ \begin{array}{l}
    dS \\ dM 
    \end{array} \right]^{4f} 
    = 
 \left[   \begin{array}{cc}
     \frac{C}{T_0} & \gamma \\ \gamma & \chi 
    \end{array} \right]^{4f}
\left[     \begin{array}{l}
    dT \\ dH 
    \end{array} \right]^{4f} 
\end{aligned},
\end{equation}
where $C^{4f} = T_0 \frac{\partial S^{4f}}{\partial T}$, $\gamma^{4f} = \frac{\partial S^{4f}}{\partial H}$, and $\chi^{4f} = \frac{\partial M^{4f}}{\partial H}$ are the \textit{intrinsic} specific heat, magnetocaloric, and isothermal magnetic susceptibility coefficients associated with the 4f subsystem respectively, and $T_0$ is the bath temperature for the 4f subsystem (provided by the lattice, $T^{lat}$ below). 

If there were no heat exchange between the 4f subsystem and the lattice, there would be no entropy change in the 4f subsystem, $dS^{4f}=0$. Thus we need to consider non-adiabatic heat exchange between the 4f subsystem  and the lattice, as well as between the lattice and the external heat bath, 
\begin{equation}\label{4f_heat_equation}
\begin{aligned} 
        q_{4f \leftarrow lat} &= T^{4f}\frac{dS^{4f} }{dt}  = 
    -\kappa_{int} (dT^{4f}-dT^{lat})  
        \end{aligned}
        \end{equation} 
where $q_{4f \leftarrow lat}$ is heat flux from lattice to the $4f$ spins and $\kappa_{int}$ is the heat link between the two. These equations need to be accompanied by equations for heat flow into the lattice, both from the 4f subsystem and from the heat bath:
\begin{equation}\label{lattice_heat_equation}
\begin{aligned} 
T^{lat}\frac{dS^{lat} }{dt} &= -  q_{4f \leftarrow lat} - q_{bath \leftarrow lat}  \\
&= \kappa_{int} (dT^{4f}-dT^{lat}) - \kappa_{ext} (dT^{lat} -dT^{bath})\\
\frac{dS^{lat}}{dt}&=\frac{C^l}{T^{lat}}\frac{dT^{lat}}{dt}
\end{aligned}
\end{equation}
where $dS^{lat}$ is the change of entropy of the lattice, and $\kappa_{ext}$ is the thermal link between the lattice and the bath. Together, Eq. \ref{4f_cross_relaxation}, \ref{4f_heat_equation} and \ref{lattice_heat_equation} describe completely the experimental setup illustrated in Fig. \ref{fig-1}(b). 

We note that all three sets of equations are {\it instantaneous}: thermodynamic relations in Eq. \ref{4f_cross_relaxation}--by its nature, and the heat flow equations Eq. \ref{4f_cross_relaxation}, \ref{4f_heat_equation}, by approximation that the heat conductance $\kappa_{int}$ and $\kappa_{ext}$ are constants, exhibiting no frequency dispersion in the experimental frequency range. We can therefore use a Fourier approach to analyze the response as a function of frequency of the driving field, $\omega$, 

\begin{equation}\label{define_ac}
\begin{aligned}
    dH(t)&= dH(\omega) e^{-i\omega t}  \\
     dM(t)&= dM(\omega) e^{-i\omega t}  \\
   dS^{4f,lat}(t)&=dS^{4f,lat}(\omega)e^{-i\omega t}  \\
   dT^{4f,lat}(t)&=dT^{4f,lat}(\omega)e^{-i\omega t}  \\
\end{aligned} 
\end{equation}
to write a system of equations for frequency components, where the frequency components on the right-hand side of the equation are all complex functions of $\omega$:
\begin{equation}
\begin{alignedat}{3}\label{system_of_equations}
   \left[ \begin{array}{l}
    dS(\omega) \\ dM(\omega) 
    \end{array} \right]^{4f} 
    = & 
 &&\left[ \begin{array}{cc}
     \frac{C}{T_0} & \gamma \\ \gamma & \chi 
    \end{array} \right]^{4f}
\left[     \begin{array}{l}
    dT(\omega) \\ dH(\omega) 
    \end{array} \right]^{4f}  \\
     \\
        -i\omega T^{4f}  dS^{4f}(\omega)   
    =& -&&\kappa_{int} (dT^{4f}(\omega)-dT^{lat}(\omega))  \\ 
-i\omega T^{lat} dS^{lat}(\omega) 
=& &&\kappa_{int} (dT^{4f}(\omega)-dT^{lat}(\omega)) \\
&-&&\kappa_{ext} (dT^{lat}(\omega)-dT^{bath}(\omega))
\end{alignedat}
\end{equation}
Note that we assume the bath holds at constant temperature $T^{bath} = T_0$, $dT^{bath}(\omega)=0$, and the temperature variation of each subsystem is much smaller than the total value (i.e. $T^{lat}\approx T_0$, $T^{4f}\approx T_0$, which is valid for considering the linear response.) Equation \ref{system_of_equations} is then a system of 4 equations for 4 unknowns ($dT^{4f}(\omega)$, $dT^{lat}(\omega)$, $dM^{4f}(\omega)$ and $dS^{4f}(\omega)$) which can be solved to find the physical quantities of interest. We will treat $dH(\omega)$ as a constant-amplitude applied AC field, and $dS^{lat}(\omega)$ is completely defined by $(C^l/T^{lat})dT^{lat}(\omega)$. Note also that $dT^{4f}(\omega)$, $dT^{lat}(\omega)$, $dM^{4f}(\omega)$ and $dS^{4f}(\omega)$ are all complex quantities, with both an amplitude and a relative phase with respect to the driving AC magnetic field.

We can now calculate the two relevant response functions, the physical quantities that will be measured in experiments, 
\begin{equation}
\begin{aligned}
\chi_{AC}(\omega) = \frac{ dM^{4f}(\omega)}{dH(\omega)}\,, \quad \text{and} \qquad 
\Gamma_{AC}(\omega) = \frac{ dT^{lat}(\omega)}{dH(\omega)}
\end{aligned}
\end{equation}
where $\chi_{AC}(\omega)$ is the effective magnetic response of the experimental setup, and $\Gamma_{AC}(\omega)$ is the effective magnetocaloric response measured through the lattice subsystem. Both of these response functions are defined via total (not partial) derivatives with respect to the driving field. Consequently, due to the internal thermal exchange process between the 4f and lattice subsystems, and the external relaxation between the lattice and the environment, both response functions differ from the intrinsic response of the magnetic ions defined by Eqn 2. 

% \onecolumngrid
% Solving for $\chi_{AC}(\omega)$ and $\Gamma_{AC}(\omega)$ in Eq. \ref{system_of_equations}, we obtain:
%  \begin{equation} \label{MCE_solution_1}
%     \begin{aligned}
%         &\chi_{AC}(\omega)=\chi^{4f}-\frac{T_0(\gamma^{4f})^2(i\omega)(\kappa_{int}+\kappa_{ext}-C^l(i\omega))}{-\kappa_{int}\kappa_{ext}+(i\omega)(C^l\kappa_{int}+C^{4f}(\kappa_{int}+\kappa_{ext}))-C^{4f}C^{l}(i\omega)^2} \\
%     \end{aligned}
% \end{equation}
%  \begin{equation} \label{MCE_solution_2}
%     \begin{aligned}
%         &\Gamma_{AC}(\omega) = \frac{-T_0 \gamma^{4f} \kappa_{int} (i\omega)}{-\kappa_{int} \kappa_{ext}+(C^{4f}+C^l)\kappa_{int} (i\omega)+C^{4f} \kappa_{ext}(i\omega)-C^{4f} C^l (i\omega)^2} \\
%     \end{aligned}
% \end{equation}

% \onecolumngrid
Solving for $\chi_{AC}(\omega)$ and $\Gamma_{AC}(\omega)$ in Eq.~\ref{system_of_equations}, we obtain:

\begin{widetext}
\begin{align}
\chi_{AC}(\omega)
&= \chi^{4f}
-\frac{T_0(\gamma^{4f})^2(i\omega)
(\kappa_{int}+\kappa_{ext}-C^l(i\omega))}
{-\kappa_{int}\kappa_{ext}
+(i\omega)(C^l\kappa_{int}+C^{4f}(\kappa_{int}+\kappa_{ext}))
-C^{4f}C^{l}(i\omega)^2}
\label{MCE_solution_1}
\\[4pt]
\Gamma_{AC}(\omega)
&=
\frac{-T_0 \gamma^{4f} \kappa_{int} (i\omega)}
{-\kappa_{int} \kappa_{ext}
+(C^{4f}+C^l)\kappa_{int} (i\omega)
+C^{4f} \kappa_{ext}(i\omega)
-C^{4f} C^l (i\omega)^2}
\label{MCE_solution_2}
\end{align}

% \twocolumngrid
\end{widetext}
The real and imaginary parts of $\chi_{AC}(\omega)$ and $\Gamma_{AC}(\omega)$ are plotted in Fig. \ref{fig-1}(c, d) for parameter values closely matching the experiment performed on YbVO$_4$. This corresponds to a scenario in which $\frac{\tau_{int}}{\tau_{ext}} > 1$, which is ideal if one intends to determine the intrinsic $\tau_{int}$ in an experiment. Solutions corresponding to a wider set of parameter values are shown in Appendix \ref{app:solution_plot}. Note that we have chosen to use a convention in which the complex numbers are formed following a positive sign: i.e.  $\chi_{AC}=\chi_{AC}'+i\chi_{AC}''$  and $\Gamma_{AC}=\Gamma_{AC}'+i\Gamma_{AC}''$.\footnote{From an experimental perspective, care must be taken to correctly establish the sign of the imaginary component of $\Gamma_{AC}$.  In practice, this will depend on the setup of the lock-in(s) receiving the MCE signal and must be carefully checked in any experiment. See Appendix \ref{app:relative_phase} for more details.} As we show below, both of these response functions are indeed observed in our measurements of \ce{YbVO4}.

Seen on a Cole-Cole plot (right hand panel of Fig \ref{fig-1}(c)), the magnetic response $\chi_{AC}(\omega)$ follows a distorted semicircular path, exclusively in one quadrant (+,+), with the external thermal relaxation determining the deviations from a perfect semi-circle at lower frequencies (right hand side of the distorted semicircle).  For smaller values of $\frac{\tau_{int}}{\tau_{ext}}$ the response evolves into two distinct semicircles (see Appendix \ref{app:solution_plot}). In contrast, the magnetocaloric response $\Gamma_{AC}(\omega)$ exhibits a full circle, inhabiting two quadrants, (+,+) and (+,-), independent of the value of $\frac{\tau_{int}}{\tau_{ext}}$ (see Appendix \ref{app:solution_plot}). 

Looking at the real and imaginary parts of $\chi_{AC}(\omega)$ and $\Gamma_{AC}(\omega)$ in Fig \ref{fig-1}(c) and (d) for the specific value of $\frac{\tau_{int}}{\tau_{ext}}$ that is used and that corresponds to the observations, the peak in the real part of $\Gamma_{AC}(\omega)$  occurs at approximately the same frequency as the maximum in the out of phase component of $\chi_{AC}(\omega)$, and occurs at a frequency that reflects the geometric mean of the thermal relaxation times $\tau_{int}$ and $\tau_{ext}$. This correspondence is more complicated for larger values of $\frac{\tau_{int}}{\tau_{ext}}$, as demonstrated in Appendix \ref{app:solution_plot}.

\begin{figure}[ht!]
	\includegraphics[width = \columnwidth]{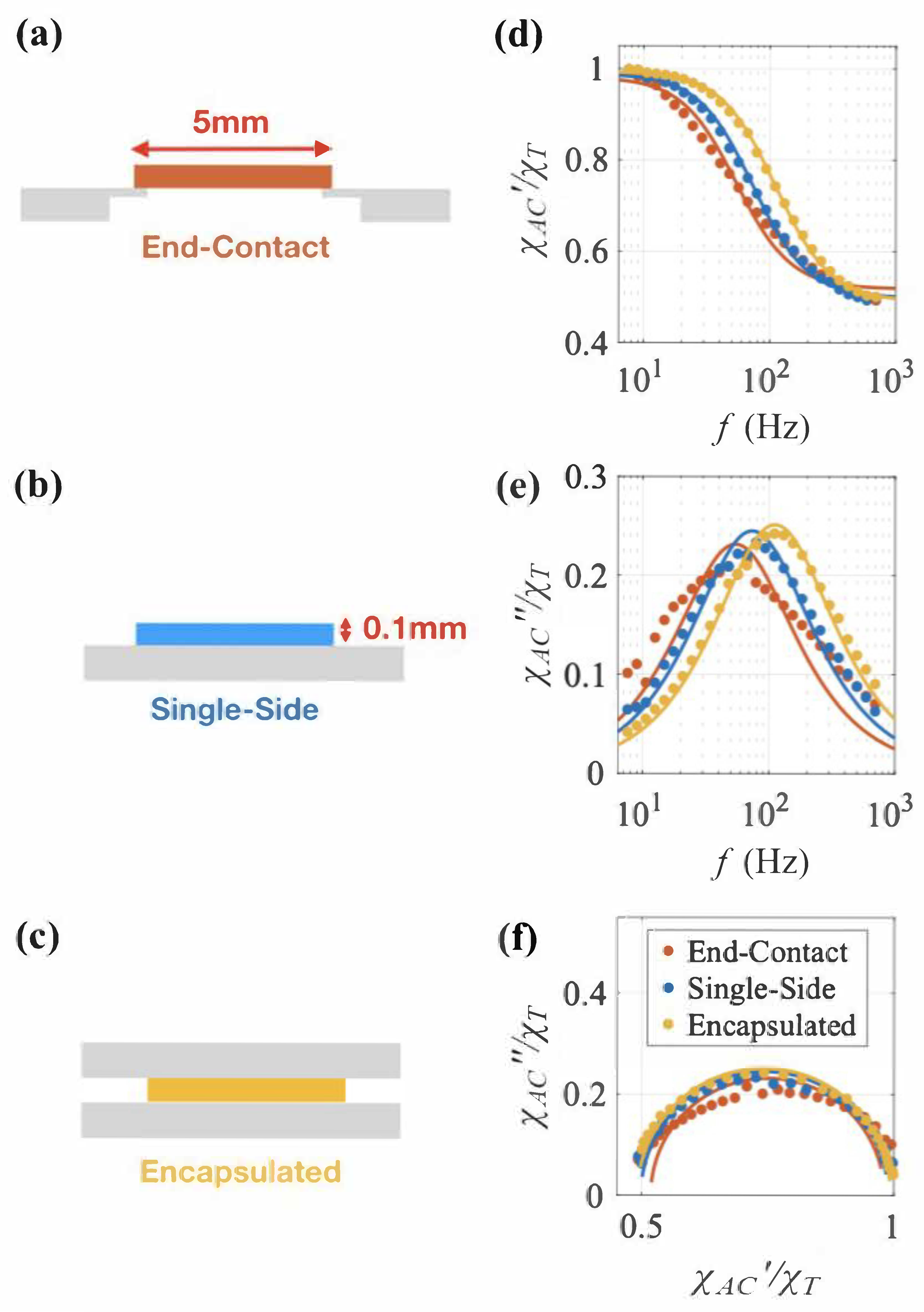}
	\caption{\label{AC_chi} Data illustrating how sample mounting configurations can affect AC magnetic relaxation. Panel (a-c) illustrates three different sample mounting configurations, described in the main text. Colored bars represent the sample that is to be measured, together with its dimensions. The same sample of \ce{YbVO4} is used for all three configurations to enable direct comparison. Grey blocks indicate quartz platforms that serve as heat baths, or at least thermal connection to a larger heat bath. The sample is oriented with the magnetic field aligned along the long c-axis of the crystal (horizontal in the schematic diagrams). AC susceptibility measurements were made at 3 K, 0.1 T, using an AC field of 3 Oe. Panels (d) and (e) show the real ($\chi_{AC}'$) and imaginary parts ($\chi_{AC}''$) normalized by $\chi_T$, which was determined from fits to the Debye model. Panel (f) shows the associated Cole-Cole plot in which frequency is an implicit variable. Data are shown for the three mounting configurations, using the same colors to differentiate the three configurations. Solid lines in (d,e,f) are fit results based on Eq. \ref{Debye}. The encapsulated configuration (yellow data points) yields data that are closest to the idealized Debye relaxation conditions. }

\end{figure}

\section{AC magnetic susceptibility}\label{sec: ac susceptibility}
Several aspects of the AC magnetic susceptibility deserve additional commentary before we proceed with a description of the experiment. 

First, as noted above, $\chi_{AC}(\omega)$ differs from the intrinsic magnetic susceptibility of the 4f subsystem $\chi^{4f}$, because the system shares energy between the two subsystems and with the thermal bath. Assuming that the intrinsic spin susceptibility $\chi^{4f}$ has no frequency dependence, the measured response $\chi_{AC}(\omega)$ obtains all of its frequency dependence through interaction with other subsystems and/or the thermal bath. In this case, the zero frequency (isothermal) value of $\chi_{AC}(\omega)=\chi^{4f}$, and $\chi_{AC}(\omega)$ fully describes the frequency dependence of the magnetic response. However, technically this does not have to be the case, and $\chi^{4f}$ can in principle have an intrinsic frequency dependence (for example, in the limit $\omega \rightarrow \infty$, the inertia of the electrons ultimately limits their ability to respond to  the driving field). For the modest frequency regimes considered here, however, such effects are irrelevant, and the intrinsic value $\chi^{4f}$, defined in Equation \ref{4f_cross_relaxation}, can be safely assumed to be independent of frequency. 

Second, we note that Eq. \ref{MCE_solution_1} naturally yields the familiar Debye relaxation \cite{Topping_2019} when considered in a limit that a single relaxation dominates. It is well known that when a single relaxation time $\tau$ governs the periodic flow of energy in a system, the dynamical susceptibility as a function of driven frequency $\omega$ is given by a Debye relaxation process \cite{Topping_2019}: $\chi(\omega)= \chi_T + \frac{1}{k} \frac{i\omega\tau}{(1-i\omega\tau)}$, where $\chi_T$ is the isothermal susceptibility (the zero frequency value), and the high-frequency value (often referred to as the adiabatic susceptibility) $\chi_S = \chi_T - \frac{1}{k}$ (i.e. $\frac{1}{k}$ is the difference between the susceptibility measured at zero frequency and infinite frequency \footnote{We use $k$ here since the expression is often derived for a damped harmonic oscillator (see for example ref \cite{Topping_2019}), in which $k$ is the spring constant. However, it should be clear for the present argument that this is simply a way to parameterize a constant that is related to the difference in the isothermal and adiabatic susceptibility.}). When the material is decoupled from the external bath (in the limit $\kappa_{ext}\rightarrow0$), Equation \ref{MCE_solution_1} reduces to: 
\begin{equation}
    \begin{aligned}  \chi_{AC}(\omega)&=\chi^{4f}-\frac{T_0(\gamma^{4f})^2}{C^l+C^{4f}}\frac{1}{1-i\omega\tau_{int}}+\frac{T_0(\gamma^{4f})^2}{C^{4f}}\frac{i\omega\tau_{int}}{1-i\omega \tau_{int}}\\
    \end{aligned}
\end{equation}
If we take the limit $C_l \rightarrow\infty$ such that the lattice acts as a thermal bath for the spins, this further reduces to the standard Debye form
\begin{equation} \label{Debye}
    \begin{aligned}  \chi_{AC}(\omega)&=\chi^{4f}+\frac{T_0(\gamma^{4f})^2}{C^{4f}}\frac{i\omega\tau_{int}}{1-i\omega \tau_{int}}\\
    \end{aligned}
\end{equation}
The isothermal, DC, limit is clearly given by $\chi^{4f}$, the intrinsic susceptibility of the 4f subsystem. Meanwhile, the adiabatic limit is given by $\chi_S=\chi_{4f}-\frac{T_0(\gamma^{4f})^2}{C^{4f}}$. An alternative, but ultimately equivalent derivation of this result is given in Appendix \ref{app:Simplified-Thermal}.

Finally, we note that the thermal model that we have developed is only applicable in the absence of significant thermal gradients. Care must be taken to closely match these conditions. This effect is explored in Section \ref{AC_magnetic_response} below, in which we describe how optimal thermal conditions can be achieved and experimentally verified.

\section{AC magnetic response of \NoCaseChange{\ce{YbVO4}}}
\label{AC_magnetic_response}

The above insights indicate that the thermal response of a material cannot, and indeed must not, be neglected in the measurement of the magnetic response at finite frequency. This motivates a more careful investigation of the cross-relaxation that occurs in AC susceptibility experiments, and that we demonstrate explicitly here for \ce{YbVO4} .

We start by measuring $\chi_{AC}(\omega)$ for a single crystal of \ce{YbVO4} mounted using different methods, with the aim of demonstrating an important point, namely that the measured value can depend on the experimental configurations that are used, even when internal relaxation processes dominate the quasi-adiabatic response. 

To illustrate the effects of extrinsic factors that indirectly affect the measured $\chi_{AC}(\omega)$ of \ce{YbVO4}, we first compare the magnetic relaxation behavior for the same crystal mounted using three different configurations, illustrated respectively in panels (a), (b) and (c) of Fig. \ref{AC_chi}. In the first case of end-contact (panel (a), red data points), the sample's extrinsic thermal contact is minimized on both ends using Low-Temperature GE Varnish and Teflon materials with relatively low thermal conductivity at low temperatures. In this configuration, the long axis of the sample exceeds the characteristic thermal length scale for all frequencies considered, and large temperature inhomogeneities can be anticipated. In the second case of the single-sided mounting condition (panel (b), blue data points), the sample has one complete face connected to the sample holder via GE Varnish. This is consistent with the standard procedure recommended by the MPMS system vendor, but still risks some degree of thermal inhomogeneity if the sample thickness is close to the characteristic thermal length scale. In the third case, the extrinsic thermal contact is maximized by encapsulating the sample between two quartz surfaces (panel (c), yellow data points), and providing the best approximation to an isothermal environment. Fig. \ref{AC_chi}(d,e) plots the measured response $\chi_{AC}(\omega)$ as a function of frequency for all three mounting configurations. 

Inspection of the data shown in Fig. \ref{AC_chi} reveals that the adiabatic (high frequency limit) and isothermal (low frequency limit) susceptibilities are independent of the extrinsic mounting configuration. Similarly, the magnitude of the excitation field does not affect the adiabatic and isothermal response (we varied the amplitude of the driven field between 1 Oe and 10 Oe, finding that this did not affect the result). Nevertheless, the way that the sample is mounted clearly affects the measured susceptibility, including the frequency at which the imaginary part of the response has its maximum value, and the overall shape of the response as a function of frequency.

To further visualize the relaxation behavior, the real part of the $\chi_{AC}(\omega)$ is plotted against its imaginary part as a Cole-Cole plot in Fig. \ref{AC_chi}(f). Under uniform magnetic and thermal conditions, and in the limit of there being a single characteristic relaxation time in the measured frequency range, the response plotted in this way should be described by a perfect semicircle with an aspect ratio of 1, where frequency is the implicit variable. As described in Section \ref{sec: heat exchange model}, including separate internal and external thermal relaxation leads to deviations from a perfect semicircle (see Appendix \ref{app:solution_plot}), but with only small deviations occurring in the low frequency regime for cases where $\frac{\tau_{int}}{\tau_{ext}} > 1$. Significantly, the data obtained from the end-contacted method (red data points) deviate from a semi-circle over the entire frequency range. In contrast, the data obtained from the single-sided and encapsulated techniques (blue and yellow data points respectively) deviate progressively less dramatically from a perfect semicircle, with the closest to semicircular behavior being obtained for the encapsulated technique. 

We emphasize that deviations from a Debeye-like response as reflected by a semi-circular shape on a Cole-Cole plot cannot be fit by the model described in Section \ref{sec: heat exchange model} and hence indicate significant thermal anisotropy. This deviation is particularly prominent for the end-contacted method (red data points). Such a mounting condition is inappropriate for determining the intrinsic susceptibility of the material and/or the associated intrinsic thermal relaxation times. Similarly, the presence of an almost perfect semicircular response for the encapsulated (yellow data points) method does not necessarily imply that there is only one relaxation time. With reference to Appendix \ref{app:solution_plot}, such an observation simply implies that $\tau_{int} > \tau_{ext}$ and/or that low enough frequencies, and/or sensitive enough measurements, are not able to distinguish the subtle deviations from perfect semicircular behavior that are present when internal and external relaxation are both present.

Importantly, we emphasize that the optimal mounting configuration is not necessarily one that minimizes thermal contact with the environment (end contacted example in this case), but rather is one that leads to the most uniform thermal conditions in the sample (fully encapsulated example in this case). Identifying the optimized thermal condition enables us to proceed with conducting quantitative analysis of the cross-relaxation affecting the magnetic and thermal response of the material in question.  We use this experimental configuration for the subsequent analysis described in Section VII.

\section{AC magnetocaloric response of \NoCaseChange{\ce{YbVO4}}}

\begin{figure}[tp]
	\includegraphics[width = \columnwidth]{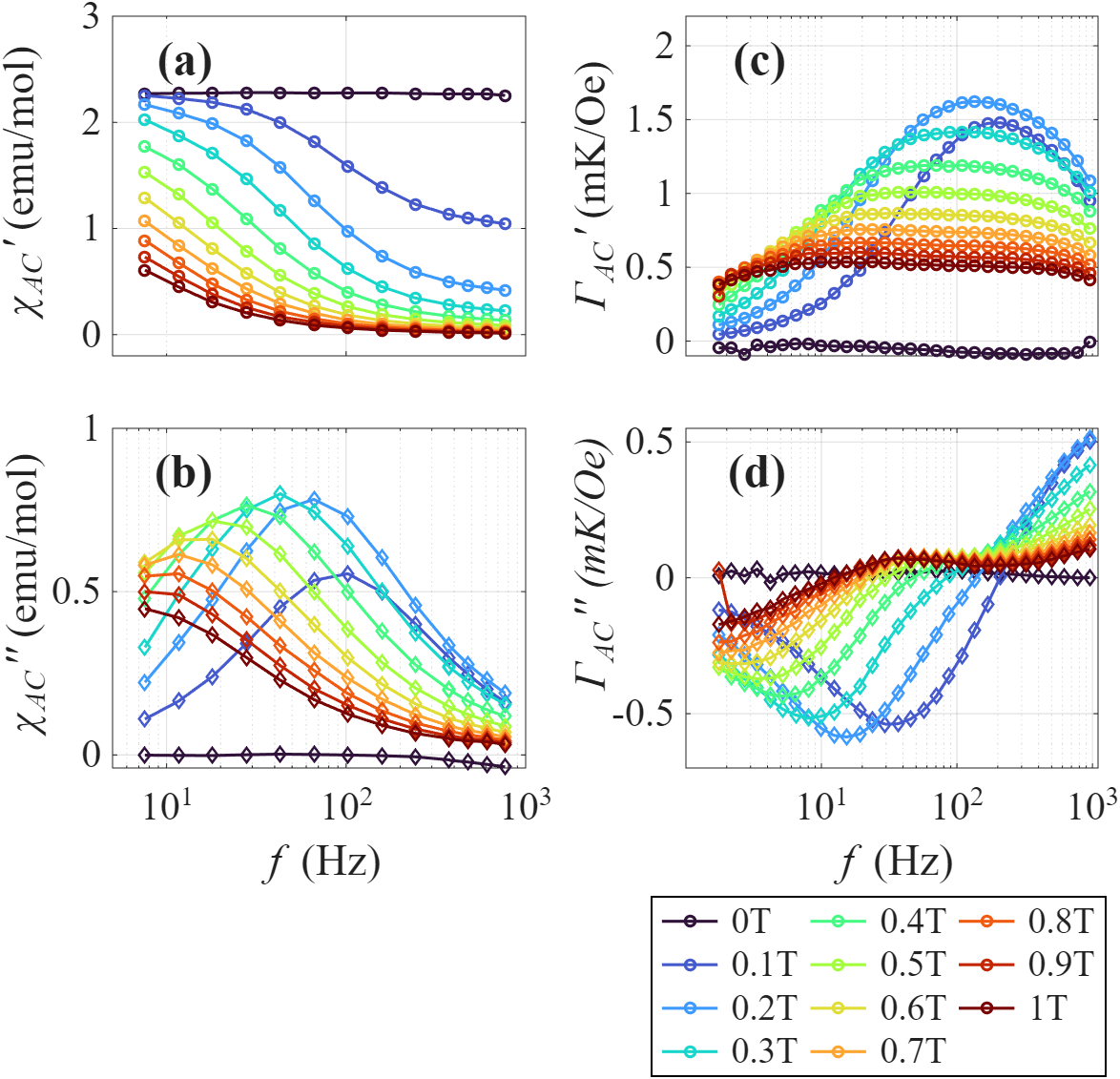}  
	\caption{\label{YbVO4_exps} Experimental results for \ce{YbVO4} showing the frequency dependence of the real (upper panels) and imaginary (lower panels) parts of the response functions $\chi_{AC}(\omega)$ (in panels (a,b)) and $\Gamma_{AC}(\omega)$ (in panels (c,d)). Data are shown for representative DC magnetic fields from 0 T to 1 T. All data were taken at a temperature of 3 K.}
\end{figure}

MCE measurements were performed following the procedures described in Section II. An AC field of 3 Oe was used. In order to ensure meaningful comparison of MCE and susceptibility measurements, the same sample, and the same mounting configuration (corresponding to the yellow configuration shown in Fig. \ref{AC_chi}), were used for both measurements. Following the discussion in Section III, the response function $\Gamma_{AC}(\omega)$ is measured, and reported in Fig. \ref{YbVO4_exps} together with $\chi_{AC}(\omega)$. In both cases, data were obtained over the same frequency range at a temperature of 3 K for magnetic fields between 0 and 1 T.

Before performing detailed fits to the data based on the thermal model, we first note three observations that are evident simply by inspecting the data shown in Fig. \ref{YbVO4_exps}. First, the data clearly follow the anticipated functional forms sketched in Fig. \ref{fig-1}(c, d). Second, there is clearly a strong field dependence to both quantities, as anticipated for \ce{YbVO4} due to Zeeman splitting of the CEF eigenstates. And third, the field-dependence is very similar for the two quantities; i.e. the maximum in the real part of $\Gamma_{AC}(\omega)$ follows the same trend as the maximum in the imaginary part of $\chi_{AC}(\omega)$. The consistency in the shift of the characteristic frequency between the two measurements implies that the same internal and extrinsic heat transfer processes are being captured in both experiments.

We note that the strong frequency dependence is not due to external relaxation. Similar measurements were performed for the closely related isotropic material \ce{GdVO4}, a material for which phonon bottleneck effects are not anticipated due to the absence of CEF effects. These measurements (see Appendix \ref{app:GdVO4_exp}) did not reveal a strong frequency dependence, even when the \ce{GdVO4} sample was held under similar experimental conditions (i.e. similar external thermal relaxation). Thus, the relaxation effects evident in Fig. \ref{YbVO4_exps} point to an intrinsic internal slow relaxation, as anticipated for \ce{YbVO4}, with the anticipated strong field dependence. Nevertheless, as the fitting procedure described below reveals, the external relaxation must still be considered in order to obtain a complete quantitative description of the experiment, which is the main point of this paper and of our analysis. Indeed, with multiple possible relaxation processes, one cannot formally identify the physical origin of the relaxation from susceptibility measurements alone (i.e. Equation \ref{Debye} would equally well describe the susceptibility if the lattice were perfectly thermally coupled to the 4f subsystem, but thermal relaxation to an external bath dominated the dynamics).\\

\section{Fits to the thermal model}

We now perform fits to measured response functions $\chi_{AC}(\omega)$ and $\Gamma_{AC}(\omega)$ shown in Fig. \ref{YbVO4_exps} to obtain the field dependence of $\tau_{int}$ and $\tau_{ext}$. 

Since the observed magnetic response is so closely approximated by a single relaxation time (i.e. the Cole-Cole plot is indistinguishable from a semicircle for the range of frequencies that we have available), we use the simplified Debye relaxation described by Eq. \ref{Debye} to extract a single relaxation time, $ \tau_{\chi}$. As we will see, this value closely approximates $\tau_{int}$ obtained from MCE measurements, consistent with the expectation that $\tau_{int} > \tau_{ext}$, and consequently can also be used in a global fit of both data sets. Both approaches were taken, such that we obtained a value $\tau_{\chi}$ from just the magnetic response, and a value for $\tau_{int}$ from a global fit to both the magnetic and caloric response. 

For the MCE data, two relaxation times are required to fit the data, corresponding to $\tau_{int}$ and $\tau_{ext}$.

To simplify the fitting procedure, we define parameters $\eta=C^l/C^{4f}$, $B=(T_0(\gamma^{4f})^2)/\kappa_{int}$, and $Q=T_0\gamma^{4f}\eta/(\kappa_{int}(1+\eta))$, where $\gamma^{4f}$ is the magnetocaloric coefficient defined in Eq \ref{4f_cross_relaxation}. Rewriting the four fitting equations after substitution of the redefined fitting parameters to Eq \ref{Debye} and \ref{MCE_solution_2} yields the following useful re-parameterizations:

\begin{equation} \label{fit_equation_chi}
    \begin{aligned}
        \chi_{AC}'(\omega)&=\chi-B+\frac{B}{1+\tau_{int}^2  \omega ^2}\\
         \chi_{AC}''(\omega)&=\frac{B\tau_{int}\omega }{1+\tau_{int}^2\omega ^2}\\
    \end{aligned}
	\end{equation}

\begin{widetext}
% \onecolumngrid
\begin{equation}\label{fit_equation_MCE}
    \begin{aligned}
        &\Gamma_{AC}'(\omega)=&\frac{-Q(1+\eta)\tau_{ext}(\tau_{ext}+\tau_{int})\omega^2}{\tau_{int}(\eta^2+(1+\eta)((1+\eta)\tau_{ext}^2+2\tau_{ext}\tau_{int}+(1+\eta)\tau_{int}^2)\omega^2+(1+\eta)^2\tau_{ext}^2\tau_{int}^2\omega^4)}\\
        &\Gamma_{AC}''(\omega)=&\frac{r\,Q\tau_{ext}\omega(\eta-(1+\eta)\tau_{ext}\tau_{int}\omega^2)}{\tau_{int}(\eta^2+(1+\eta)((1+\eta)\tau_{ext}^2+2\tau_{ext}\tau_{int}+(1+\eta)\tau_{int}^2)\omega^2+(1+\eta)^2\tau_{ext}^2\tau_{int}^2\omega^4)}\\
    \end{aligned}
\end{equation}
% \twocolumngrid
\end{widetext}

\begin{figure}[b]
    \centering
    \includegraphics[width=0.8\columnwidth]{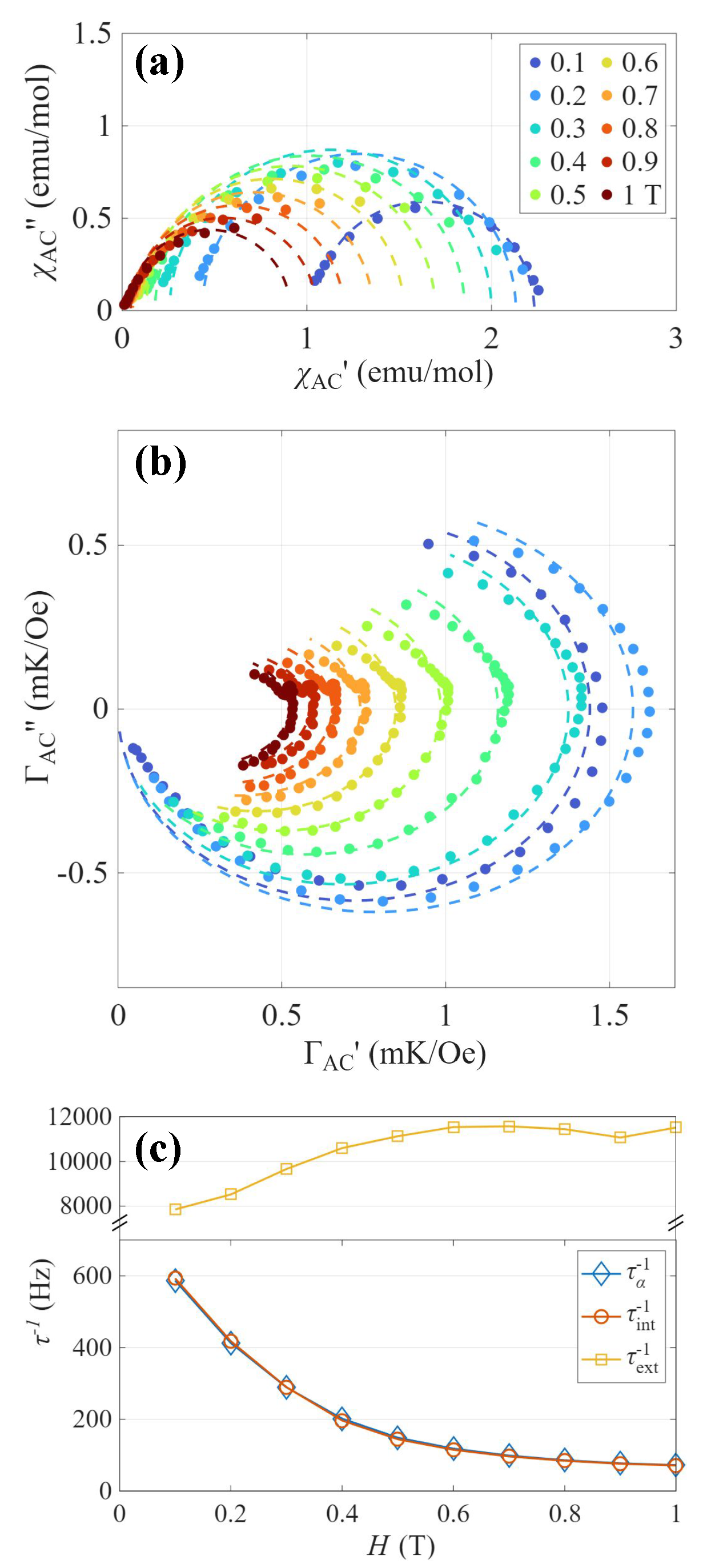}
    \caption{Fitting of the two measured response functions $\chi_{AC}(\omega)$ and $\Gamma_{AC}(\omega)$ for \ce{YbVO4} using Eqn.s \ref{fit_equation_chi} and \ref{fit_equation_MCE}. All data were measured at 3K. (a) A Cole-Cole plot showing the imaginary ($\chi_{AC}''$) against the real ($\chi_{AC}'$) part of the dynamical susceptibility. (b) A similar plot for the AC MCE response, $\Gamma_{AC}(\omega)$. (c) The characteristic relaxation times ($\tau_\chi$, $\tau_{int}$ and $\tau_{ext}$) obtained from the fits, as described in the main text. See Appendix \ref{app:best fit params} for other fitted parameters.}
    \label{fitresult}
\end{figure}

Global fits of both data sets to Eq. \ref{fit_equation_chi} and \ref{fit_equation_MCE} yield the fits shown by dashed lines in Fig. \ref{fitresult} panels (a) and (b). The fits accurately capture the observed response for both $\chi_{AC}(\omega)$ and $\Gamma_{AC}(\omega)$, implying that we have identified the appropriate thermal model, and that fit parameters can be meaningfully extracted.

Nevertheless, inspection of Fig. \ref{fitresult}(b) reveals a slightly non-circular MCE response on the Cole-Cole plot. Such an effect can be caused by external factors not captured by the simplified model here, such as the heat exchange between the material and the thermometer (see Appendix \ref{app:Thermometer}). To account for this, we include a prefactor $r$ to $\Gamma_{AC}''(\omega)$ in Eqn. \ref{fit_equation_MCE}, which is treated as empirical `non-circular' parameter in the associated fits.

The field dependence of the two time constants that are extracted from these fits is shown in Fig. \ref{fitresult} (c). The red data points show the results for $\tau_{int}$ when both $\chi_{AC}(\omega)$ and $\Gamma_{AC}(\omega)$ are simultaneously fit by the full thermal model involving two time constants. Yellow data points show the associated $\tau_{ext}$ values from those same fits to the full thermal model. In order to confirm that the combination fitting method is valid, we also fit the measured $\chi_{AC}(\omega)$ separately using only Eq. \ref{fit_equation_chi}.
Blue data points in Fig. \ref{fitresult}(c) show the result from fitting $\chi_{AC}(\omega)$ using only a single-time constant denoted as $\tau_{\chi}$. We find that the fitted $\tau_{ext}$ relaxes much faster than $\tau_{int}$, implying that $\tau_{int}/\tau_{ext}>1$ is satisfied here, as previously mentioned in Section \ref{sec: heat exchange model}. This is also consistent with the close agreement between $\tau_{int}$ and $\tau_\chi$, which demonstrates that the two experiments capture the same energy transfer processes, and that the intrinsic internal time constant dominates in this case. 

The focus of this paper is on developing the experimental technique of AC MCE and the associated thermal model. Nevertheless, we comment briefly on quantitative aspects of the experimental result specific to YbVO$_4$. First, the derived values for $\tau_{int}$ are consistent with typical values found for other similar materials with strong magnetic anisotropy \cite{Orbach1961}. More interestingly, as we show in Appendix \ref{app:best fit params}, we find that the field dependence follows an exponential relationship, indicating that the Zeeman splitting of the CEF eigenstates plays a significant role in determining $\tau_{int}$. (Indeed, following Orbach's analysis \cite{Orbach1961}, this indicates that the Debye temperature of YbVO$_4$ is larger than the CEF splitting, as indicated schematically in Fig. 1.) And finally, relative to the strong field-dependence of $\tau_{int}$, $\tau_{ext}$ exhibits a much smaller variation with field. The physical origin of this small field-dependence presumably reflects a field-induced change in $\kappa_{ext}$, but this would need to be checked by separate measurements of the field-dependence of the thermal conductivity of YbVO$_4$.

Despite the quantitative success of the thermal model in describing the observed response functions, nevertheless there are subtle deviations from the anticipated behavior. In particular, the MCE data shown in Fig. \ref{fitresult}(b) deviate slightly from a perfect circle (captured by the prefactor $r$ in Eqn. \ref{fit_equation_MCE}) and also reveal a subtle 'kink', that is most evident for higher field data. Empirically, this kink is found to be sensitive to the way in which the thermal sensor is attached to the sample (see Appendix \ref{app:Thermometer}). Since this has only a modest effect for the present experiment, for clarity and simplicity we explicitly neglect including the thermal sensor in the thermal model that we have developed and that is captured by Eqn.s \ref{MCE_solution_2} and \ref{fit_equation_MCE}. However, as discussed in Appendix \ref{app:Thermometer}, it is relatively simple to include this as an additional discrete element in the thermal model if needed (i.e. if heat flow between the lattice and the thermal sensor is a significant factor).

We close this section by commenting on the fit to the magnetic response, $\chi_{AC}(\omega)$. For cases where the external relaxation time is considerably smaller than the internal one (such as this case) $\chi_{AC}(\omega)$ follows an almost semi-circular form on a Cole-Cole plot (see Appendix \ref{app:solution_plot}). Consequently, Eq. \ref{MCE_solution_1} cannot be used to separately determine the internal and external time constants, and instead, as we have done, one must use the simplified Eq. \ref{Debye}. Generally speaking, although Debye-like relaxation behavior obtained from the magnetic response can be fitted with a simplified (single relaxation time) model, the interpretation of the time constant that is obtained from such a fit could be incomplete or even misleading unless the corresponding AC MCE measurement identifies the relation between the time constants. In particular, taken on its own, it is not even evident if the obtained time constant is due to internal or external relaxation. The MCE result must be considered in parallel in order to reveal the complete thermal exchange process.

\section{Discussion}
It is instructive to consider the relation between the thermal analysis that we present above to a generalized linear response theory \cite{Landau, Onsager}. When multiple variables interact with each other in a circuit, the overall relaxation behavior is characterized by a set of thermodynamic conjugate variables, which refer to pairs of ``force" ($f_\omega$) and ``displacement" ($x_i(\omega)$) that respond directly to each other. A change in one variable directly affects the other. The application of ``forces" results in a corresponding ``displacement" that characterizes the linear response function $\alpha_i(\omega)$:

\begin{equation} \label{conjugate_responce}
    \begin{aligned}
        x_i(\omega)=\alpha_i(\omega) f_\omega
    \end{aligned}
	\end{equation}

The most commonly existing pair of conjugate variables is temperature and entropy $(T, S)$. When the thermodynamic state of a closed system remains unchanged by applied fields, the thermal-susceptibility is uniquely defined by the heat capacity because $ dS=(C_p/T) dT$. Consequently, the dynamical response is completely determined by the rate of thermal conductance, affected by an external heat source. If, however, the system develops a caloric response, it must imply that another pair of conjugate variables exists in at least one of the circuit components. For example, the measurement results above contain 3 pairs of conjugate variables, which are $(H, M)$, $(T^{4f}, S^{4f})$ and $(T^{lat}, S^{lat})$. Consequently, the full relaxation behavior can be completely determined by a linear response matrix:

\[
\begin{blockarray}{cccc}
& M & S^{4f} & S^{lat} \\
\begin{block}{c(ccc)}
  H & \alpha_{11} & \alpha_{12} & \alpha_{13} \\
  T^{4f} & \alpha_{21} & \alpha_{22} & \alpha_{23} \\
  T^{lat\,} & \alpha_{31} & \alpha_{32} & \alpha_{33} \\
\end{block}
\end{blockarray}
 \]

The diagonal terms represent the susceptibility response. $\alpha_{11}$ is the AC magnetic susceptibility ($\chi_{AC}(\omega)$ in Eq. \ref{MCE_solution_1}), $\alpha_{22}$ accounts for the thermal susceptibility of the spin affected by the magnetic field, and $\alpha_{33}$ accounts for thermal susceptibility of the lattice. From an experimental perspective, we note that any conventional dynamical measurement only obtains the frequency response function of the diagonal matrix element. Although all matrix elements must satisfy the relation of $\alpha_{i,i+j}\alpha_{i+j,i}=\alpha_{i,i}\alpha_{i+j,i+j}$, the matrix cannot be uniquely defined by dynamical susceptibility only. Furthermore, the term $\alpha_{33}$ is special in the sense that it is a single positive real constant $C^{l}/T_0$ instead of a complex response function. The AC caloric effect function $\Gamma_{AC}(\omega)$ described in this paper can be associated with $\alpha_{13}$ by the relation $\Gamma_{AC}(\omega)=\alpha_{13}/\alpha_{33}$. 

The matrix response shown above provides an illustrative example that any dynamical behavior of a non-instantaneous magnetic response is most likely to be the complex result of both internal structure and external conditions. Hence, for systems that are composed of multiple degrees of freedom, the measured $\chi_{AC}(\omega)$ on its own is insufficient to reflect the complete dynamical process. The discretized thermal model based on spin-lattice relaxation underscores the importance of measuring the AC caloric response as a reciprocal observation to provide complementary information to any susceptibility analysis. 

Finally, we note that a natural extension of the above analysis is to also consider elastocaloric responses, where the `force' and `displacement' terms literally correspond to components of the stress and strain tensor, the diagonal terms in the linear response matrix correspond to terms in the elastic stiffness tensor, and the off-diagonal terms correspond to the elastocaloric response. 
 
\section{Summary}

We have developed an experimental method to measure the AC MCE, and have introduced a discretized thermal analog circuit approach that fully describes the cross-relaxation between magnetization and temperature in the frequency domain in the presence of an AC magnetic field excitation. This approach permits a full understanding of the frequency dependence of the AC susceptibility $\chi_{AC}(\omega)$ and the AC MCE $\Gamma_{AC}(\omega)$ through the presence of both extrinsic and intrinsic relaxation processes. 

We demonstrated the technique and the associated fitting methods using \ce{YbVO4}, a material for which there are slow magnetization dynamics at low temperatures arising from a phonon-bottleneck effect, with a strong field dependence.

The magnetic dynamics of a wide variety of materials are often inferred from just susceptibility measurements, and analysed using a Debye, or closely related, model. Associated slow dynamics can be attributed to various physical effects. While these analyses are well-motivated, they are necessarily incomplete if cross-relaxation is neglected, potentially missing or mis-characterizing important new material-specific insights. The new caloric approach developed in this paper could provide additional evidence about these internal relaxation phenomena, while also providing a grounded description of the accompanying thermal relaxation effects associated with extrinsic effects. We expect that the approach we have outlined here will prove helpful in future studies of magnetic dynamics in a wide range of materials.

\section{Acknowledgements}
We thank Matthias S. Ikeda for fruitful discussions and original insights on the AC caloric measurement techniques. Low-temperature measurements performed at Stanford University were supported by the Air Force Office of Scientific Research Award FA9550-24-1-0357, using a cryostat acquired with award FA9550-22-1-0084. Crystal-growth experiments were supported by the Gordon and Betty Moore Foundation Emergent Phenomena in Quantum Systems Initiative Grant GBMF9068. Work at Los Alamos National Laboratory is supported by the NSF through DMR-1644779 and DMR-2128556, the U.S. Department of Energy, and the DOE/BES “Science of 100 T” grant. M.P.Z. was also partially supported by a National Science Foundation Graduate Research Fellowship under grant number DGE-1656518. L.Y. acknowledges support from the Marvin Chodorow Postdoctoral Fellowship at the Department of Applied Physics, Stanford University. 

\begin{appendices}

\section{\NoCaseChange{Experimental set up of the AC MCE measurement} \label{app:mce_exp}}

We employ a RuO$_x$ thin film resistor as the thermal sensor. The sensor is mounted directly on the sample and connected to a Wheatstone bridge, comprising three other identical RuO$_x$ resistors (Fig. \ref{fig-7}(a)). The resistance of the sensor is measured using an AC current, with a drive frequency $\omega_c \sim$ 3137 Hz. The magnetic field is modulated at a fixed frequency $\omega_m$, with values between 0.1 and 1000 Hz. Using a dual-frequency lock-in technique, we demodulate the signal to obtain the amplitude and relative phase of the thermally-driven changes in resistance of the RuO$_x$ sensor.  

 The circuit diagram is shown in (Fig. \ref{fig-7}(b)). We use a Stanford Research SR860 lock-in amplifier operating in dual mode to measure the bridge voltage $V_b$ (proportional to the \textit{change} in resistance of the RuO$_{x}$ sensor, from which the amplitude and phase of the temperature oscillations of the thermal sensor are deduced), and another SR860 operating in external mode to measure the actual resistance of the $V_S$ (from which the average temperature of the sample is deduced). The amplified driven coil signal gives an additional isolated output via the Electronic Module of the MPMS System. The voltage output with the internal reference frequency of the first lock-in was amplified into a current signal via a Stanford Research CS580 Voltage Controlled Current Source and also gave the reference signal input of the second lock-in. The Electron Transport Option (ETO) probe provided by the MPMS accessory kits was applied to enable transport measurement. 
 
 Several aspects of the circuit and the operation of the dual mode lock in amplifier deserve a more detailed commentary. 

\subsubsection{Obtaining the amplitude of temperature oscillations}
In an AC MCE measurement, we convert the amplitude signal $V_{SR860}=\sqrt{X_{out}^2+Y_{out}^2}$ to the actual temperature signal recorded by the sensor. In the circuit diagram shown in Fig. \ref{fig-7}(b), a change of resistance $\Delta R$ is detected by $R_x$. The bridge voltage $V_b$ is defined as the peak-to-peak value, therefore it is 2 times the value of $V_{sr860}$ reported from the electronics \cite{SR860}:

    \begin{equation} \label{lock_in_singal_conversion}
    \begin{aligned}
   V_b&=2V_{SR860}\\
   I_b&=2I_R
    \end{aligned}
	\end{equation}

 According to the property of the Wheatstone bridge, we have:
    \begin{equation} \label{bridge_voltage_conversion}
    \begin{aligned}
   V_b&=-\frac{V(R+\Delta R)}{2R}+\frac{VR}{2R}\\
   &=-\frac{V\Delta R}{2R}
    \end{aligned}
	\end{equation}

By knowing the characteristic $R-T$ curve of the thermoresistor, the value of the temperature derivative can be obtained from a polynomial curve fitting method. Therefore, the value of $\frac{dR}{dT}$ is well-defined at each temperature. We calculate the change of temperature $\Delta T$ as:

    \begin{equation} \label{bridge_voltage_conversion}
    \begin{aligned}
   \Delta T&=-\Delta R/(\frac{dR}{dT})\\
   &=-\frac{4V_{SR860}}{I_b}/(\frac{dR}{dT})
    \end{aligned}
	\end{equation}

Finally, the response function $\Gamma_{AC}(\omega)$ is experimentally obtained by $\Gamma_{AC}(\omega)=\Delta T(\omega)/\Delta H$.

\subsubsection{Determining the relative phase of the temperature oscillations}\label{app:relative_phase}

It is worth noting that the phase measured in the dual-frequency mode of SR860 depends on the choice of the internal reference frequency. Specifically, swapping the internal and external references flips the sign of the phase. As an example, consider a modulated signal in the form 
\begin{equation}
V_\text{sig}(t)=\text{sin}(\omega_ct)(1+\delta V\,\text{sin}(\omega_mt+\phi_m))
\end{equation}
with the carrier frequency $f_c=2\pi\omega_c$ and the modulation frequency $\omega_m$.

If the internal frequency is set to $\omega_c$ and an external reference passed in at $f_m=2\pi\omega_m$, then the SR860 multiplies the signal by $X_\text{LIA}=\sqrt{2}\,\text{sin}((\omega_c-\omega_m)t)$ in Channel X and by $Y_\text{LIA}=\sqrt{2}\,\text{cos}((\omega_c-\omega_m)t)$ in Channel Y \cite{SR860}, leading to the outputs:

\begin{equation}
    \begin{aligned}
    X_c&=\sqrt{2}\,\text{sin}((\omega_c-\omega_m)t)\,\text{sin}(\omega_ct)(1+\delta V\,\text{sin}(\omega_mt+\phi_m))\\
        &=\frac{\delta V}{2\sqrt{2}}\text{sin}(\phi_m)+...\\
    Y_c&=\sqrt{2}\,\text{cos}((\omega_c-\omega_m)t)\,\text{sin}(\omega_ct)(1+\delta V\,\text{sin}(\omega_mt+\phi_m))\\
        &=\frac{\delta V}{2\sqrt{2}}\text{cos}(\phi_m)+...
    \end{aligned}
\end{equation}
where ``..." include the oscillating terms that are filtered out.
On the other hand, if the internal reference is at $f_m$ and the external reference at $f_c$, then $X_\text{LIA}$ acquires a minus sign while $Y_\text{LIA}$ remains the same. Therefore, the outputs become:
\begin{equation}
    \begin{aligned}
    X_m&=\sqrt{2}\,\text{sin}((\omega_m-\omega_c)t)\,V_\text{sig}(t)=-\frac{\delta V}{2\sqrt{2}}\text{sin}(\phi_m)+...\\&=-X_c\\
    Y_m&=\sqrt{2}\,\text{cos}((\omega_m-\omega_c)t)\,V_\text{sig}(t)=\frac{\delta V}{2\sqrt{2}}\text{cos}(\phi_m)+...\\&=Y_c
    \end{aligned}
\end{equation}
and the measured phase is complex conjugate to the previous case.

\begin{figure}[tp]
	\includegraphics[width = \columnwidth]{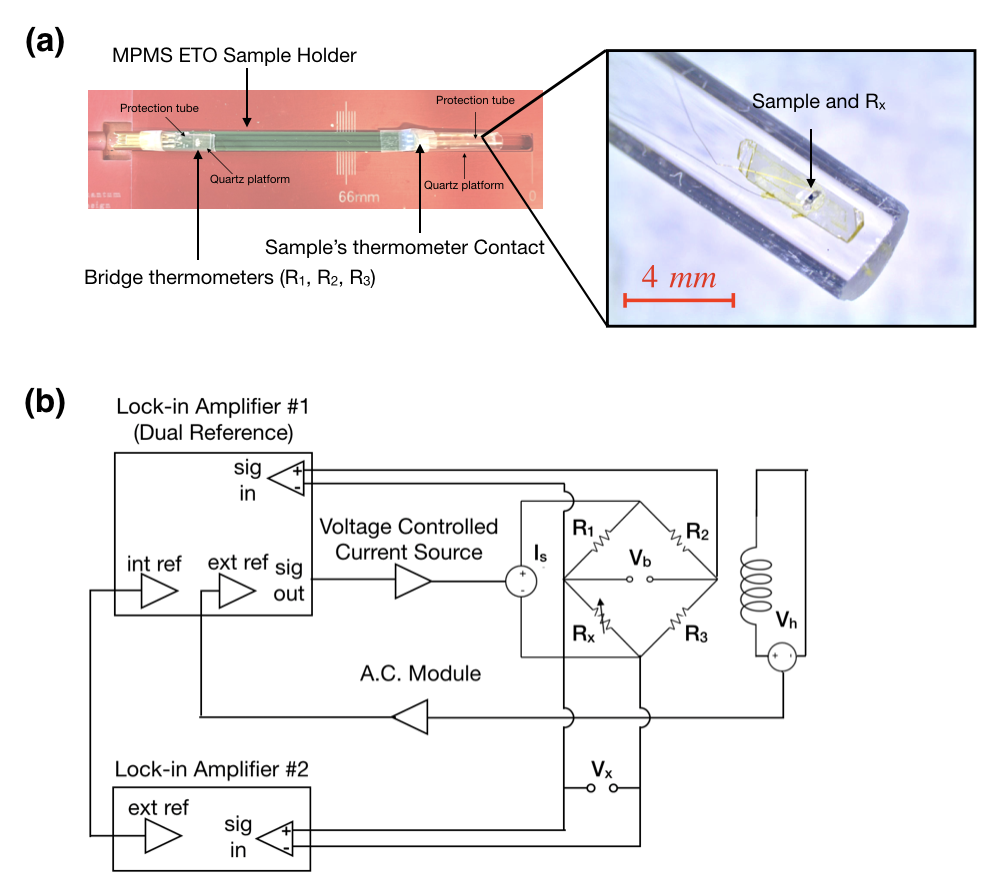}  
	\caption{\label{fig-7} (a) Photograph of an AC MCE probe assembly, with inset showing a magnified image of the thermal sensor ($R_x$; the small black chip in the image) mounted on the crystal of \ce{YbVO4} (yellow color prism) which is mounted on a flattened section of a thin quartz rod (clear). The sample is located at the end of the quartz rod, which is attached to the ETO probe head. Before measurement, the center of the sample is aligned with the center of the applied magnetic coils following standard procedures. (b) Circuit diagram of the AC MCE measurement. Here, $R_x$ represents the thermal resistor. $R_1$, $R_2$, and $R_3$ are bridge resistors. The raw temperature oscillation signal is obtained from the bridge voltage $V_b$ via a lock-in amplifier in a dual mode with its internal reference, and external reference from the magnetic coil ($V_h$). A second lock-in amplifier measures $V_x$ to obtain the temperature profile of the thermal resistors. A Voltage Controlled Current Source converts the $1\ \text{V}$ voltage output of the lock-in amplifier to an excitation current of $100 \ \mu\text{A}$ to the Wheatstone Resistor Bridge.}
 
\end{figure}

In the case where $f_\text{int}=f_m$, outputs $X_m$ and $Y_m$ are rotated by $90^\circ$ compared to the typical measurements locked into a single frequency. In the other case where $f_\text{int}=f_c$ (e.g. Fig. \ref{fig-7}), outputs $X_c$ and $Y_c$ swap $X$ and $Y$ channels, hence equivalent to an extra conjugation after the rotation. 
We have verified this reference-dependence of phase in DUAL mode with a daisy-chain experiment: in place of the Dual mode lock-in (LIA\#1 in Fig. \ref{fig-7}), we demodulated the signal with a lock-in at $\omega_{c}$ first, and then fed its X channel output to a second lock-in at $\omega_m$. Swapping the X and Y outputs form the second lock-in gave us back the measurements done in Dual mode with $f_{int}=f_c$, as expected.

\section{\NoCaseChange{CEF eigenstates and spin-lattice relaxation in \ce{YbVO4}} \label{app:CEF}}

The CEF levels are eigenstates of the Hamiltonian:
\begin{equation} \label{CEF_Hamiltonian}
\begin{aligned}
	H_{CEF}=B_2^0 O_2^0+B_4^0 O_4^0+B_4^4 O_4^4+B_6^0 O_6^0+B_6^4 O_6^4\\
 \end{aligned}
\end{equation}
where $B_n^m$ are the CEF parameters defined by the \ce{RVO4} lattice for Yb$^{3+}$ ions in $D_{2d}$ point symmetry, and $O_n^m$ are the standard Steven operators. The irreducible representations of the ground-state, first and second excited states calculated for the \ce{Yb^{3+}} ion with these crystal field parameters \cite{NIPKO1997} are:

\begin{equation} \label{CEF_Eignestates}
\begin{aligned}
	\pm\Gamma_1^7&=0.901\ket{\pm\frac{7}{2}}-0.433\ket{\mp\frac{1}{2}}, E=0 \: meV\\
 \pm\Gamma_1^6&=\pm0.886\ket{\pm\frac{3}{2}}\mp0.464\ket{\mp\frac{5}{2}}, E=7.2 \: meV\\
 \pm\Gamma_2^7&=0.433\ket{\pm\frac{7}{2}}+0.901\ket{\mp\frac{1}{2}}, E=34.8 \: meV\\
 \end{aligned}
\end{equation}

Momentary deviations from $D_{2d}$ point symmetry, caused by phonons, allow for transitions between these erstwhile eigenstates. Phonon modes investigated by room temperature Raman spectra \cite{Santos2007, SANTOS2007_2} and inelastic neutron scattering \cite{NIPKO1997} that break the $D_{2d}$ symmetry have been reported. Introducing the additional Stevens operators $\Braket{\Gamma| O_m^n| \Gamma}$ for $m=2,4,6$ and $n=0,2,4,6$ and calculating the associated matrix elements, we find that the extended CEF parameters allow transitions between ($+\Gamma_1^7,-\Gamma_1^6$) and ($-\Gamma_1^7,+\Gamma_1^6$). There is, however, no allowed transition between/within the ground state doublet: $\Braket{\Gamma_1^7| O_m^n| -\Gamma_1^7}=0$ (this is ultimately a consequence of these being Kramers doublets, with a minimal degeneracy of 2 in the absence of breaking time reversal symmetry). Thus, the spin-lattice relaxation must occur via an indirect process, and hence the transition rate is strongly affected by thermal population of the relevant phonon modes.

\section{\NoCaseChange{AC magnetic susceptibility and AC MCE of \ce{GdVO4}}}
\label{app:GdVO4_exp}
\ce{GdVO4} has the same crystal structure as \ce{YbVO4}. The absence of CEF effects for the Gd ion means that the Hunds rule ground state remains unsplit in zero field. The material undergoes an antiferromagnetic transition at 2.5 K, followed by a spin-flop phase down to 1.37 K \cite{Mangum1972}. It is paramagnetic at 3 K. 

\begin{figure}[t!]
	\includegraphics[width = \columnwidth]{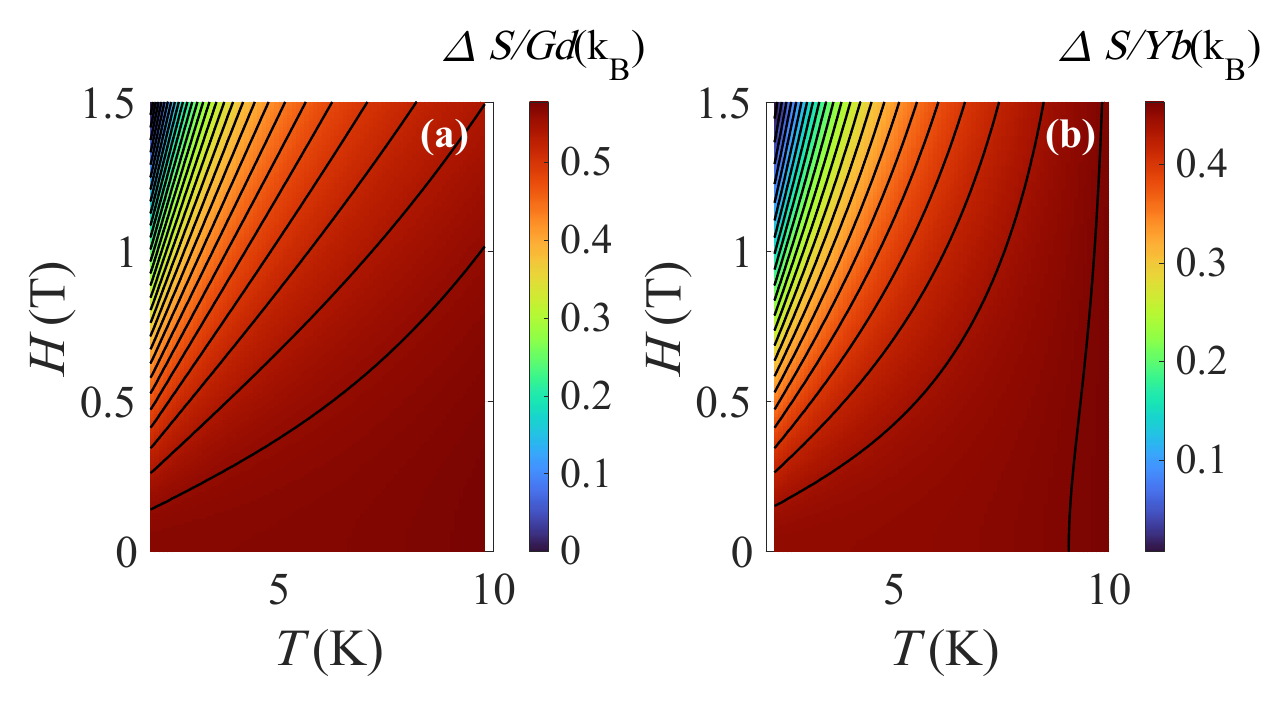}
	\caption{\label{entropy} The calculated entropy change of (a) \ce{GdVO4} and (b) \ce{YbVO4} with respect to the entropy value of 2 K and 0 T, including the entropy of the lattice and the $4f$ contribution. The black lines represent isentropic contours.}
\end{figure}

Both \ce{YbVO4} and \ce{GdVO4} exhibit a large magnetocaloric effect at low temperatures. The physical origin in both cases is of course the strong variation of the internal entropy as a function of field and temperature, due to Zeeman splitting of the magnetic states. Fig. \ref{entropy} illustrates the calculated entropy landscapes for both compounds, including the phonon background (for clarity and simplicity we neglect the magnetic order in \ce{GdVO4}). Adiabatic changes in magnetic field will yield comparably large changes in temperature for the two materials.  

\begin{figure}[t!]
	\includegraphics[width = \columnwidth]{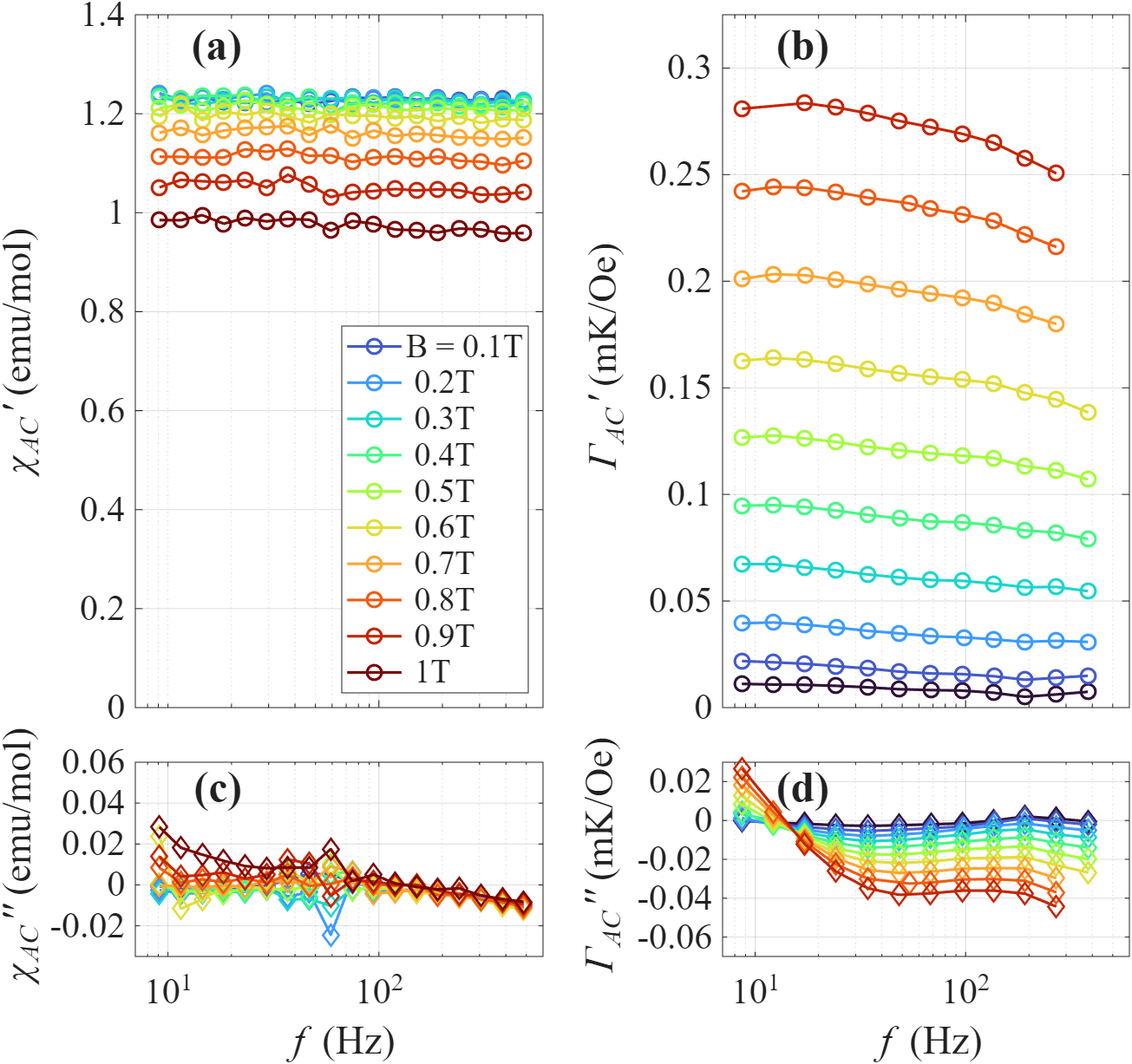}  
	\caption{\label{GdVO4_exps} Real and imaginary parts of the magnetic and thermal response functions $\chi_{AC}(\omega)$ (panels a and b) and $\Gamma_{AC}(\omega)$ (panels c and d) of \ce{GdVO4} as a function of frequency. Data were all taken at 3K, for DC fields varying from 0 T to 0.18 T. The same sample, with the same mounting configuration, was used for both measurements.}
\end{figure}

Whereas \ce{YbVO4} exhibits slow relaxation dynamics at low temperatures, no such relaxation effects are observed in \ce{GdVO4} because of the absence of any significant phonon bottleneck effects. In Fig. \ref{GdVO4_exps}, we show the measured $\chi_{AC}(\omega)$ and $\Gamma_{AC}(\omega)$ at 3 K. Notably, no significant frequency dependence was observed, and the amplitude of the out-of-phase component remained comparable to the background noise for frequencies below 100 Hz. Thus \ce{GdVO4} provides a useful counterpoint in which there are no internal relaxation effects in the frequency range measured, and extrinsic time constants associated with both measurements exceed the maximum accessible frequency of the AC experiments.

\section{\NoCaseChange{Simplified Thermal Exchange Model} \label{app:Simplified-Thermal}}

In this section, we further explain the origin of the simplified expression for  $\chi_{AC}(\omega)$ discussed in Section \ref{sec: ac susceptibility}. 

We consider a scenario in which the lattice and the heat bath in Fig. \ref{fig-1}(b) have good thermal contact, hence, the two components are considered as a single environment relative to the 4f spin subsystem. By assuming the magnetocaloric effect heats the sample uniformly in ideal thermal conditions (i.e., neglecting the kind of extrinsic mounting effects described in Appendix \ref{app:Thermometer}), the heat transfer between the spin subsystem and a constant bath is described by a thermal transfer equation:
\begin{equation} \label{simplified_thermal_model}
    \begin{aligned}
         \left[ \begin{array}{l}
    dS(\omega) \\ dM(\omega) 
    \end{array} \right]^{4f} 
    = & 
 \left[   \begin{array}{cc}
     \frac{C}{T} & \gamma \\ \gamma & \chi 
    \end{array} \right]^{4f}
\left[     \begin{array}{l}
    dT(\omega) \\ dH(\omega) 
    \end{array} \right]^{4f}  \\
     \\
        -i\omega T^{4f}  dS^{4f}(\omega)   
    =& -\kappa_{int} (dT^{4f}(\omega)-dT^{bat})  \\
    \end{aligned}
	\end{equation}
    
Holding $dT^{bat}=0$ and solving for Eq. \ref{simplified_thermal_model}, we obtain:

\begin{equation} \label{simplified_thermal_model_solution}
    \begin{aligned}
    \chi_{AC}(\omega) = \frac{ dM^{4f}(\omega)}{dH(\omega)}=\chi^{4f}-\frac{T_0(\gamma^{4f})^2(-i\omega)}{C^{4f}(-i\omega)+\kappa_{int}}\\
    \end{aligned}
	\end{equation}

The solution here is the same as the solution of $\chi_{AC}(\omega)$ solved from Eq. \ref{system_of_equations}, which we obtained by assuming that the effective $\kappa_{ext}$ goes to zero and the effective $C^l$ is much greater than $C^{4f}$ (which leads to Eq. \ref{Debye}). Good extrinsic thermal contact in Eq. \ref{system_of_equations} implies a large value of $\kappa_{ext}$, but these two models are actually the same. Intuitively, good thermal contact with the bath (large $\kappa_{ext}$) is thermally equivalent to a single, infinitely large lattice (large $C_l$)—isolated (small $\kappa_{ext}$) or otherwise. Mathematically, this can be justified with the fact that Eq. \ref{MCE_solution_1} is the same in the limit of $C_l \rightarrow \infty$ and the limit of $\kappa_{ext} \rightarrow \infty$. Therefore, the approximation/limits affecting $C^l$ and $\kappa_{ext}$ in the main text, and the simplified thermal exchange model outlined here, are equivalent in explaining the Debye-like relaxation behavior observed in $\chi_{AC}(\omega)$. Such ideas of an extended lattice bath cannot be considered in the AC MCE, because the thermometer measures a local region of the sample.

\section{\NoCaseChange{Effect of the thermal sensor on the measured response} 
\label{app:Thermometer}}

The heat exchange model in the main text overlooks the effect of the thermal sensor. This was justified because ideally the sensor only has a marginal impact on the thermal exchange between the major subsystems/components, and our goal has been to reveal the essential physics necessary to describe the major heat flows in the physical system. 

Here, we develop a generalized description of the system that treats the temperature of the sample $T^{s}(\omega)$ and of the thermometer $T^{\theta}(\omega)$ as independent values. We consider the simplest scenario, in which a thermal sensor is attached to a part of a sample with thermal conductivity $\kappa_{\theta}$. If the sample and sensor have a good thermal contact, the associated (third) time constant $\tau_{\theta} = \frac{C^lC^\theta}{\kappa_{\theta}(C^l+C^\theta)} \simeq C^\theta/\kappa_{\theta}$ will be much smaller than $\tau_{int}$ and $\tau_{ext}$, and the heat flow will be essentially instantaneous relative to the important timescales being probed by the experiment. More generally, heat flow between the sample and sensor is governed by the equation:

\begin{equation} \label{eqn_heat_exchange}
    \begin{aligned}
    q_{\theta \leftarrow sample}&=-T^{\theta}\frac{dS^{\theta}(t)}{dt}\\
    &=-\kappa_{\theta}(dT^{\theta}(t)-dT^{sample}(t)).\\
    \end{aligned}
	\end{equation}
    
Following the same principles developed in the main text, we can also write the equation in its frequency components:

\begin{equation} \label{eqn_heat_exchange_2}
    \begin{aligned}
   -i\omega dS^{\theta}(\omega)T^{\theta}&=-\kappa_{\theta}(dT^{\theta}(\omega)-dT^{sample}(\omega))\\
   dS^{\theta}(\omega)&=\frac{C^{\theta}}{T_0}dT^{\theta}(\omega).
    \end{aligned}
\end{equation}

Eq. \ref{eqn_heat_exchange_2} can be solved together with Eq. \ref{system_of_equations} to include the heat flow between the lattice and the sensor to extend the thermal model. We can then obtain expressions for the sensor temperature $\Gamma_{\theta}(\omega)$.

We note that for a well-designed experiment the response function $\Gamma_{\theta}(\omega)$ is very close to that of the material $\Gamma_{lattice}(\omega)$ ($\Gamma_{AC}(\omega)$ in the main text). Heat exchange with the thermal sensor also affects the magnetic response, modifying the expression for $\chi_{AC}(\omega)$ if those data are taken while the thermal sensor is attached. The modified expressions are given by:
\begin{equation} \label{MCE_solution_thrm_1}
    \begin{aligned}
        &\chi_{AC}(\omega)=\chi^{4f}-\frac{T_0(\gamma^{4f})^2(i\omega)((\kappa_{int}+\kappa_{ext})\kappa_\theta-C^lC^\theta\omega^2-iN\omega)}{-\kappa_{int} \kappa_{ext}\kappa_\theta+\omega^2D_1+i\omega D_2} \\
        &\Gamma_{\theta}(\omega) = \frac{-T_0 \gamma^{4f} \kappa_{int}\kappa_\theta (i\omega)}{-\kappa_{int} \kappa_{ext}\kappa_\theta+\omega^2D_1+i\omega D_2}\,,\\
        &\text{with}\\
        &\quad N=C^l\kappa_\theta+C^\theta(\kappa_{int}+\kappa_{ext}+\kappa_\theta)\,,\\
        &\quad D_1 = C^{4f} C^l\kappa_\theta+C^\theta C^l\kappa_{int}+C^{4f}C^\theta(\kappa_{int}+\kappa_{ext}+\kappa_\theta)\,,\\
        &\quad D_2 =(C^{4f}\kappa_{int}+C^l\kappa_{int}+C^{4f}\kappa_{ext})\kappa_\theta+C^\theta\kappa_{int}(\kappa_{ext}+\kappa_\theta)\\
        &\quad\quad-C^{4f}C^lC^\theta\omega^2
    \end{aligned}
\end{equation}

whereas the lattice temperature is
\begin{equation} \label{MCE_solution_thrm_3}
    \Gamma_{lattice}(\omega) = \frac{-T_0 \gamma^{4f} \kappa_{int}(i\omega)(\kappa_\theta-i C^\theta\omega)}{-\kappa_{int} \kappa_{ext}\kappa_\theta+\omega^2D_1+i\omega D_2}
\end{equation}

\begin{figure}[ht!]
    \includegraphics[width = \columnwidth]{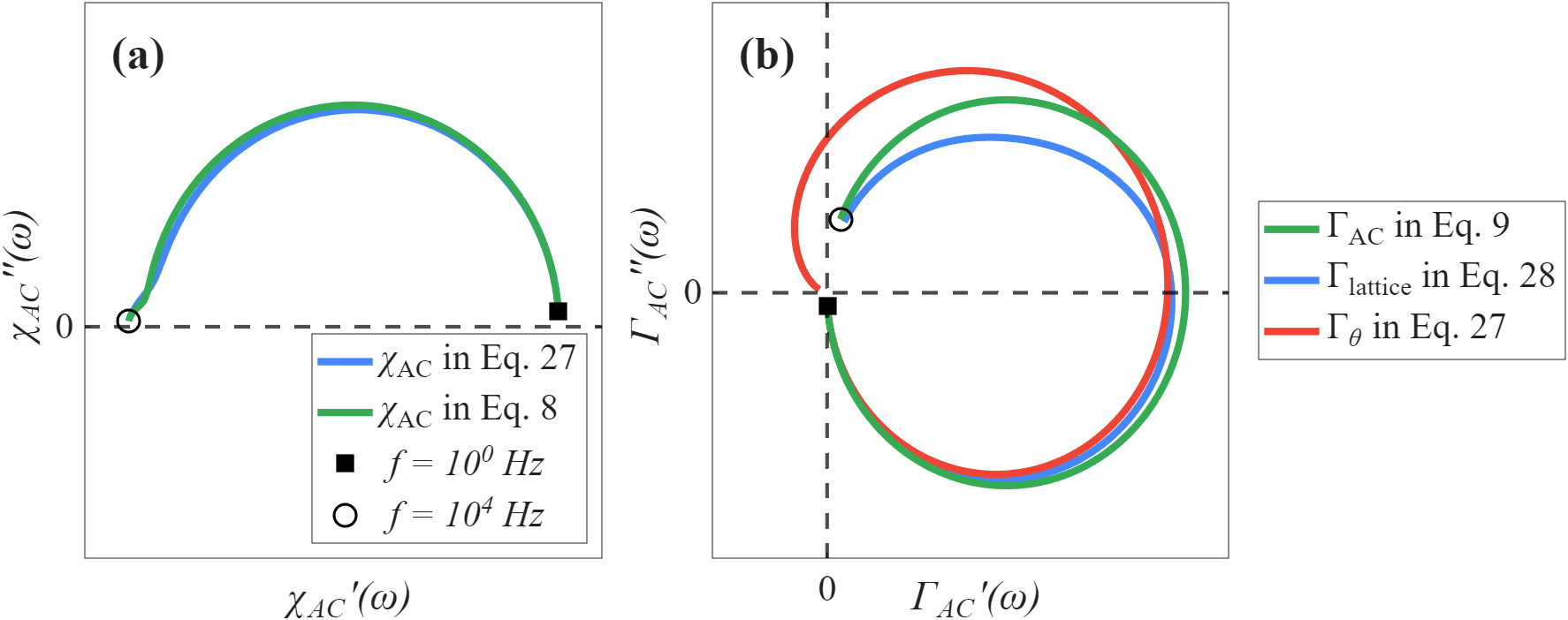}
	\caption{\label{fig-wThrm} Calculated Cole-Cole plots of the thermal model including a thermometer with finite heat capacity ($C^\theta$) and thermal conductivity ($\kappa_\theta$). Panel (a) compares the magnetic response solved with and without the thermometer, and panel (b) plots the magnetocaloric response seen in the lattice ($\Gamma_{lattice}(\omega)$) and in the thermometer ($\Gamma_{\theta}(\omega)$).
    Parameters used here are $\eta=C^l/C^{4f}=0.1$, $C^l/C^\theta=2$, $\tau_{int}/\tau_{ext}=0.5$, $\tau_{int}/\tau_\theta=0.1$.}
\end{figure}

A comparison of the solutions is illustrated in Fig. \ref{fig-wThrm}. The green and blue lines in Panel (a) plot the magnetic response $dM/dH$, which does not change significantly if we assume a reasonable choice of thermometer, namely smaller thermal mass and comparable thermal connection compared to the rest of the setup. Meanwhile, the Cole-cole plot of the magnetocaloric response is distorted from the circular behavior in the simpler model. While the thermometer and the lattice temperatures follow the simpler solution of lattice response closely at the low frequency limit, the three curves diverge at higher frequencies as the thermometer progressively decouple from the sample. At high enough frequencies, the phase of $\Gamma_{\theta}(\omega)$ goes beyond $\pi/2$, leading to an opposite change in temperature between the lattice and the thermometer.

The heat flow in the real experiment can further differ from the idealized behavior described above. In particular, since the size of the thermal sensor is typically much smaller than the sample, at sufficiently high frequencies it cannot exchange heat evenly with the entirety of the sample. This consideration has a negligible effect when the sensor is vanishingly small and in good thermal contact with the sample. However, when the thermometer has a limited rate of thermal exchange defined by either geometry or contact materials, the mounting geometry can introduce inhomogeneous thermal gradients, and hence the measured thermal response cannot be properly described by the idealized discrete thermal model. In practice, this means that care must be taken that the thermal sensor (including wiring and associated heat flow) does not overly perturb the overall heat flow so much as to obscure the MCE oscillations that are the focus of the measurement.

\begin{figure}[tp]
	\includegraphics[width = \columnwidth]{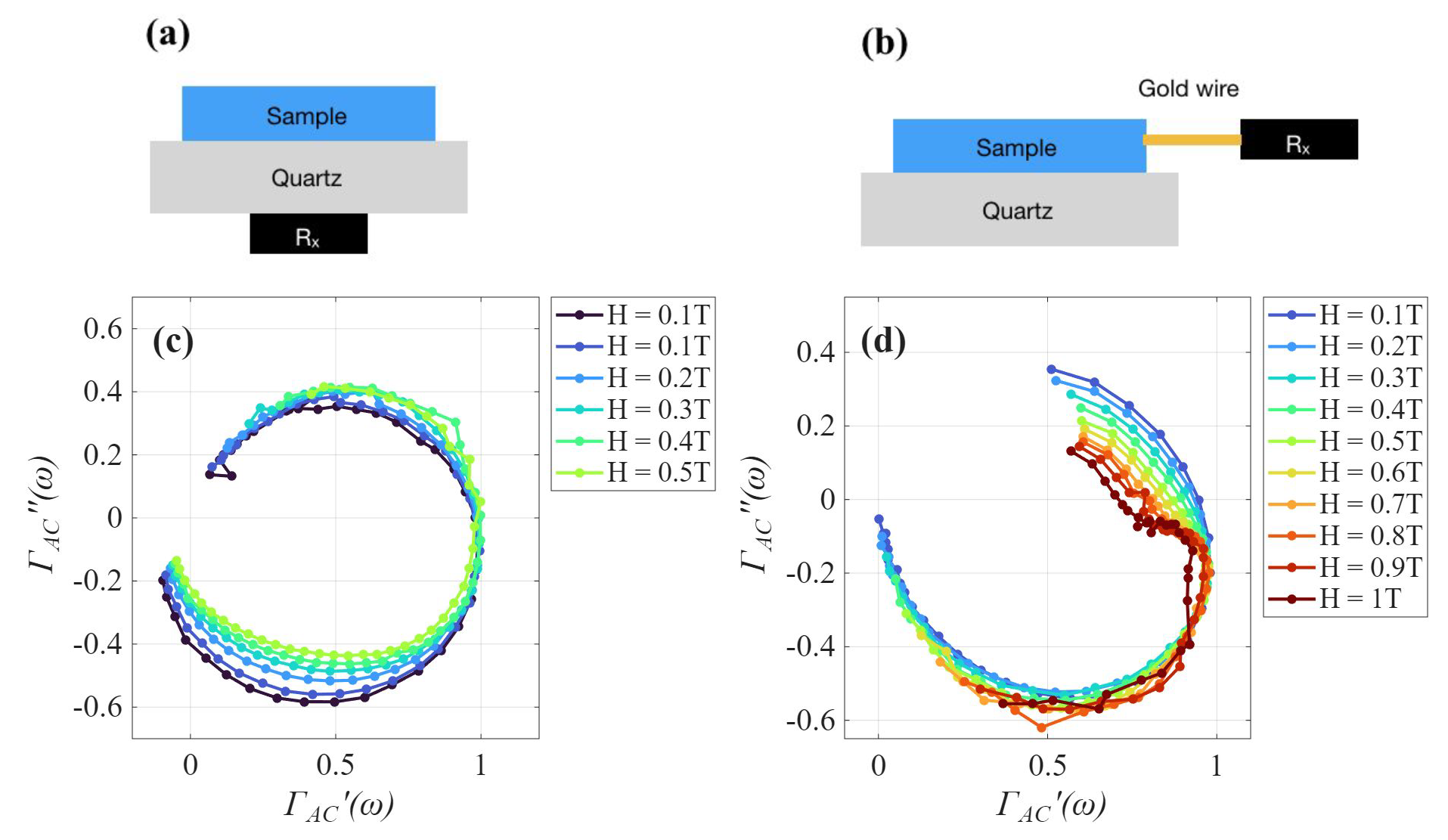}  
	\caption{\label{extrinsic_mce} (a,b) Schematic diagrams illustrating two additional methods to mount a sample for AC MCE measurements, as examples of sub-optimal mounting conditions which limit the thermal conduction rate between the sample (blue bar) and the thermal sensor (R$_x$). In (a) the thermometer and sample are separated by the quartz component that the sample is mounted on, and in (b) they are separated by a 0.05 mm diameter gold wire which serves as a heat pipe. (c) The Cole-Cole plot of $\beta/\beta_{max}$ of the sample mounted following the method shown in (a). Real and imaginary parts of $\beta$ have been normalized by the maximal value $\beta_{max}$ for each curve. (d) The Cole-Cole plot of the measurement result for the sample mounted following the method shown in (b). Both measurements were for YbVO$_4$ at a temperature of 3 K.}
 
\end{figure}

To experimentally demonstrate the extrinsic effects that the thermal sensor can introduce, we designed two additional experiments in which the thermal sensor is deliberately connected indirectly to the sample of YbVO$_4$. The two experiments are illustrated in in panels (a) and (b) in Fig. \ref{extrinsic_mce}. In the first of these, the thermal sensor is attached to the back side of the quartz plate to which the sample is affixed. In the second, the thermal sensor is attached to the sample via a thin gold wire (i.e. a heat pipe). The results are displayed as Cole-Cole plots in panels (c) and (d) of Fig. \ref{extrinsic_mce}.

When the sensor and the sample are separated by a single quartz component, the resulting Cole-Cole plot deviates from a circular shape for lower frequencies (Fig. \ref{extrinsic_mce}(c)). In particular, the Cole-Cole plot now reveals a response in \textit{three} quadrants (including negative values of $\Gamma'$ in Quadrant II). This is distinct from the case in which the thermal sensor is neglected, and for which the response is always $\Gamma' > 0$ (i.e. only in Quadrants I and IV). This behavior is described by Eqn. \ref{MCE_solution_thrm_1}, as illustrated in Fig. \ref{fig-wThrm}), and is indicative of an additional slow thermal time constant associated with relaxation between the sample and the thermal sensor. 

In contrast, when the sample and sensor are separated by a thin gold wire, deviations from the circular Cole-Cole plot are evident at higher frequencies (Fig. \ref{extrinsic_mce}(d)). This behavior is again evident of a third time constant, but with a different characteristic frequency.  

In summary, the thermal sensor can affect the observed response function. A well-designed experiment has negligible effect, and is evident from a circular Cole-Cole response. The effect of the thermal sensor can be quantitatively accounted for so long as thermal gradients are minimized.

\section{\NoCaseChange{Additional plots of the solutions of the thermal exchange model for a wider range of parameters} \label{app:solution_plot}}

\begin{figure}
	\includegraphics[width = 0.8\columnwidth]{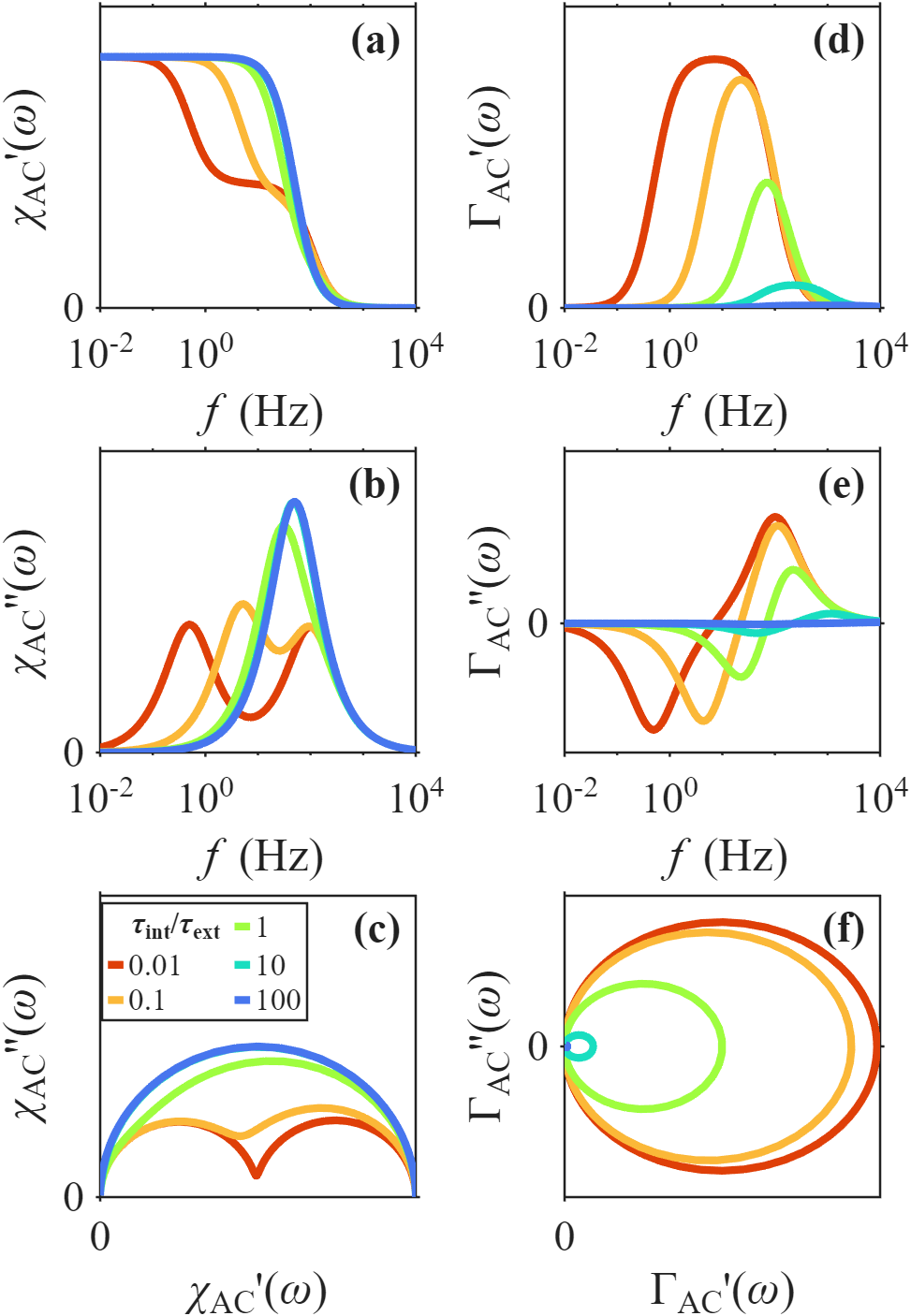}  
	\caption{\label{more_solutions} More solutions of $\chi_{AC}(\omega)$ and $\Gamma_{AC}(\omega)$ based on Eq. \ref{MCE_solution_1} and \ref{MCE_solution_2}. The plot was made by setting $T_0$, $\gamma^{4f}$, $C^{4f}$, $C^l$ and $\kappa_{int}$ as fixed value and only change the value of $\kappa_{ext}$, so that the ratio of the two time constants $\tau_{int}=(C^lC^{4f})/(\kappa_{int}(C^l+C^{4f}))$ and $\tau_{ext}=C^l/\kappa_{ext}$ can be compared when they are either close or well-separated from each other. Here we pick $C^l=C^{4f}$ for simplicity. Viewed on a normalized Cole-Cole plot (panel (c)), the magnetic response $\chi_{AC}(\omega)$ exhibits two clear semicircles for small values of $\frac{\tau_{int}}{\tau_{ext}}$, merging into a single semicircle for large values. In contrast, the thermal response $\Gamma_{AC}(\omega)$ is a circle on a Cole-Cole plot (panel (f)) for all values of $\frac{\tau_{int}}{\tau_{ext}}$. }
 
\end{figure}

The plots in Fig. \ref{fig-1} (c) and (d) illustrate a specific case: $\frac{\tau_{int}}{\tau_{ext}} > 1$ (a necessary requirement for a good measurement of the MCE response). 
To illustrate wider solutions of the thermal model described by Eq. \ref{MCE_solution_1} and \ref{MCE_solution_2}, we plot here the real and imaginary parts of the two response functions for a wider range of $\frac{\tau_{int}}{\tau_{ext}}$, obtained by changing $\kappa_{ext}$ and fixing other parameters constant. As $\frac{\tau_{int}}{\tau_{ext}}$ is reduced, a second semicircle feature emerges in the Cole-Cole plot of $\chi_{AC}(\omega)$, which is manifested as an extra plateau in $\chi_{AC}'(\omega)$ and an extra peak in $\chi_{AC}''(\omega)$. In contrast, $\Gamma_{AC}(\omega)$ remains circular on the associated Cole-Cole plot. As shown in Fig. \ref{more_solutions} (d) and (e), $\Gamma_{AC}'$ and $\Gamma_{AC}''$ grow at the same time when $\tau_{int}/\tau_{ext}$ decreases. Consequently, the radius of $\Gamma_{AC}(\omega)$ on the Cole-Cole plot (panel (f)) grows proportionally.
As described in Appendix \ref{app:Thermometer}, deviations from this perfect circle are anticipated and observed when heat exchange with the thermal sensor becomes a significant factor. 

\section{\NoCaseChange{Best-fit parameters obtained from fitting to Equations \ref{fit_equation_chi} and \ref{fit_equation_MCE}} 
\label{app:best fit params}}

\begin{figure}[h]
    \centering
    \includegraphics[width=0.8\linewidth]{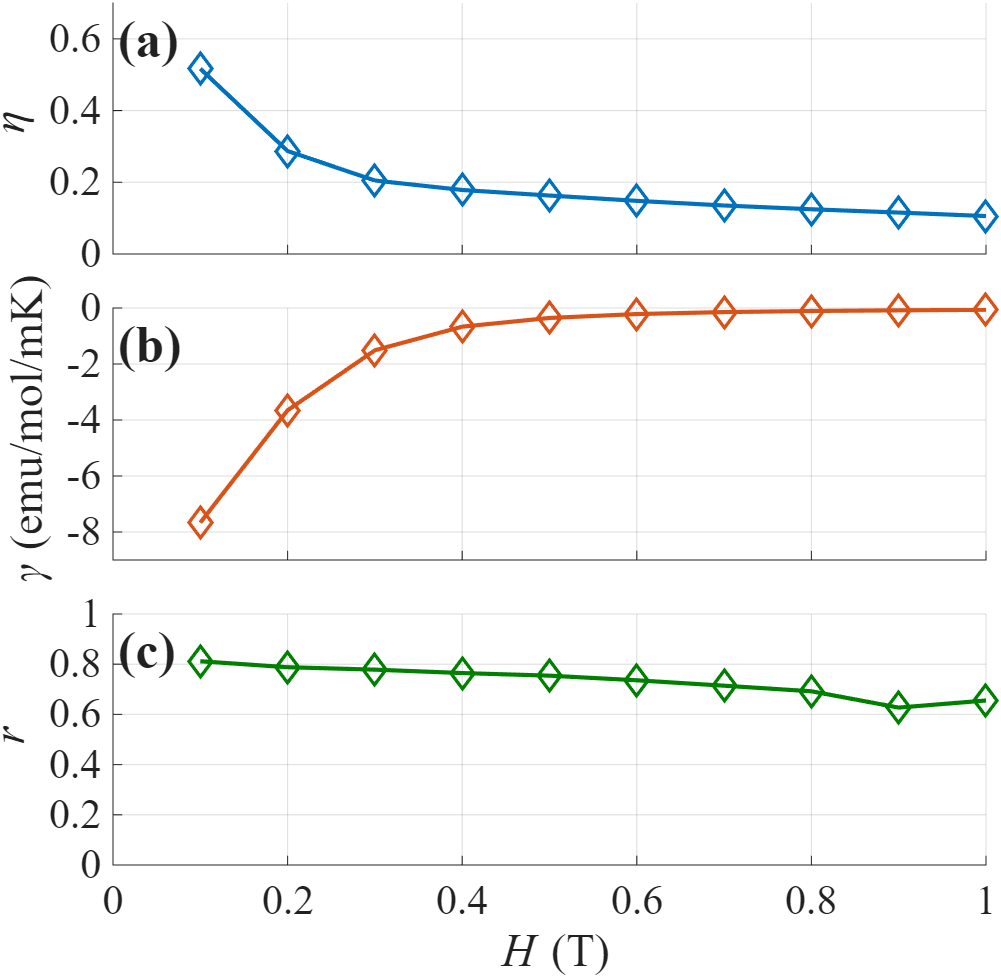}
    \caption{\label{fig:fit_params}Best fit parameters from fitting experimental data to Eqn. \ref{fit_equation_chi} and \ref{fit_equation_MCE}, including the heat capacity ratios $\eta=C^l/C^{4f}$ in (a), the magnetocaloric coefficient $\gamma^{4f}=\partial S^{4f}/\partial H|_T$ in (b), and the empirical stretching factor $r$ in (c). All parameters here are obtained by fitting to AC susceptibility and AC MCE data simultaneously, taken at different magnetic fields at 3 K.
    }
\end{figure}

In Fig. \ref{fig:fit_params}, we plot the remaining fit parameters corresponding to the dashed lines in Fig. \ref{fitresult} as a function of magnetic field. 

The empirical stretching factor $r$ has a zero-field value close to 0.8 and exhibits only a very slight field dependence. Its proximity to 1 and small field-dependence imply that the simplified heat exchange model that we use to describe this system, which uses just two time constants, captures the essential physical processes remarkably accurately. See Appendix \ref{app:Thermometer} for possible factors that contribute to $r$ deviating from unity. 

In contrast, the parameters $\eta=C^l/C^{4f}$ and $\gamma^{4f}=\partial S^{4f}/\partial H|_T$ both reveal a large field dependence. This behavior is expected given the field-induced changes in the material properties. In particular, when the magnetic field increases, $C^l$ stays constant at a fixed temperature, but $C^{4f}$ increases due to the field-induced splitting of the CEF doublet (i.e. a field-induced Schottky anomaly). Hence, $\eta$ decreases with increasing field. Similarly, as the spins are more aligned at higher field, the induced entropy change decreases, and hence the magnitude of $\gamma^{4f}$ decreases with increasing field. 

\begin{figure}
    \centering
    \includegraphics[width=0.9\linewidth]{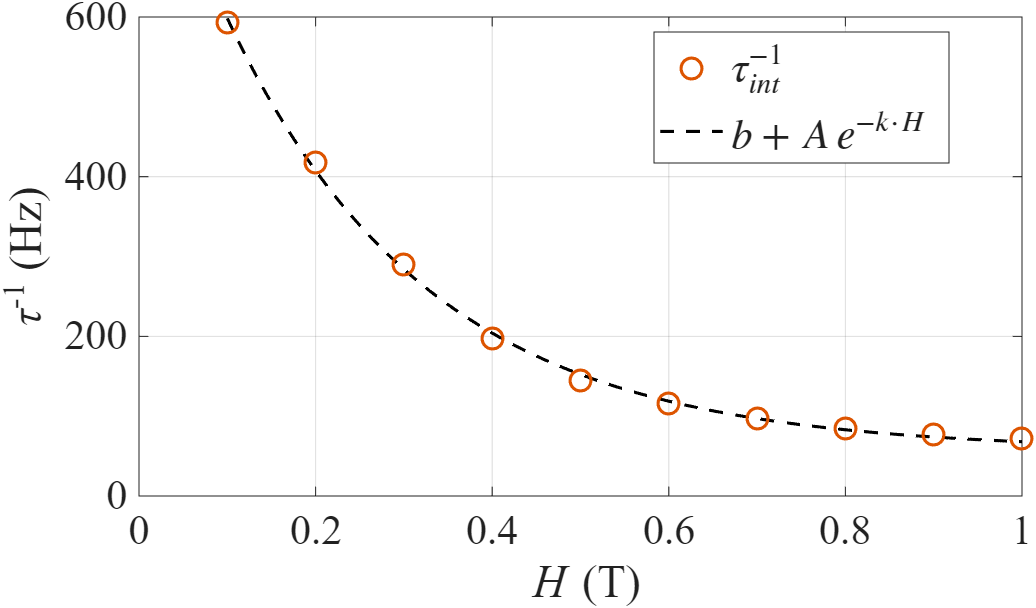}
    \caption{Exponential fit of the field dependence of $\tau_{int}^{-1}$. Red circles are the fitted spin-lattice relaxation time (same as those in Fig. \ref{fitresult}). The dashed black line plots the best-fit to $b+A\,e^{-k\cdot H}$, with parameters are $b=57.2 \text{ Hz, }A=835.6\text{ Hz, }k=4.3\text{ T}^{-1}$.}
    \label{fig:fit tau exp}
\end{figure}

Finally, we comment that the spin-lattice relaxation time exhibits an exponential behavior as a function of the magnetic field, as illustrated in Fig. \ref{fig:fit tau exp}. This is consistent with the resonance relaxation process when $k_B\Theta_D>\Delta_q$ (as elaborated by Orbach \cite{Orbach1961}), where $\Delta_q$ is the transition energy that phonons resonate with. If the energy gap between the ground state doublet and the first excited doublet at zero field is $\Delta_0$, then as the doublets split due to the Zeeman effect, the energy difference also changes linearly to the magnetic field. Therefore, we expect \[1/\tau\propto e^{-(\Delta_0+g_{\text{eff}}\mu_BB)/(k_BT)}\propto e^{-g_{\text{eff}}\mu_BB/(k_BT)}\] at low temperatures. Fitting the results at 3 Kelvin, we extracted $g_\text{eff}=k/(\mu_B/(k_BT))=19.4$.

\end{appendices}

\newpage
\bibliography{reference}

@article{Topping_2019,
doi = {10.1088/1361-648X/aaed96},
url = {https://dx.doi.org/10.1088/1361-648X/aaed96},
year = {2018},
month = {nov},
publisher = {IOP Publishing},
volume = {31},
number = {1},
pages = {013001},
author = {C V Topping and S J Blundell},
title = {A.C. susceptibility as a probe of low-frequency magnetic dynamics},
journal = {Journal of Physics: Condensed Matter},
}

@article{Coca2014,
author = {G{\'{o}}mez-Coca, Silvia and Urtizberea, Ainhoa and Cremades, Eduard and Alonso, Pablo J and Cam{\'{o}}n, Agust{\'{i}}n and Ruiz, Eliseo and Luis, Fernando},
doi = {10.1038/ncomms5300},
issn = {2041-1723},
journal = {Nature Communications},
number = {1},
pages = {4300},
title = {{Origin of slow magnetic relaxation in Kramers ions with non-uniaxial anisotropy}},
url = {https://doi.org/10.1038/ncomms5300},
volume = {5},
year = {2014}
}

@article{Kohama2010,
    author = {Kohama, Yoshimitsu and Marcenat, Christophe and Klein, Thierry and Jaime, Marcelo},
    title = {AC measurement of heat capacity and magnetocaloric effect for pulsed magnetic fields},
    journal = {Review of Scientific Instruments},
    volume = {81},
    number = {10},
    pages = {104902},
    year = {2010},
    month = {10},
    issn = {0034-6748},
    doi = {10.1063/1.3475155},
    url = {https://doi.org/10.1063/1.3475155}
}

@article{Radhakrishna1981,
title = {Neutron diffraction study of antiferromagnetism in YbVO4},
journal = {Journal of Magnetism and Magnetic Materials},
volume = {23},
number = {3},
pages = {254-256},
year = {1981},
issn = {0304-8853},
doi = {https://doi.org/10.1016/0304-8853(81)90044-5},
url = {https://www.sciencedirect.com/science/article/pii/0304885381900445},
author = {P. Radhakrishna and J. Hammann and P. Pari}
}

@article{feigelson1968flux,
  title={Flux growth of type RVO4 rare-earth vanadate crystals},
  author={Feigelson, Robert},
  journal={Journal of the American Ceramic Society},
  volume={51},
  number={9},
  pages={538--539},
  year={1968},
  publisher={Wiley Online Library}
}

@article{smith1974flux,
  title={Flux growth of rare earth vanadates and phosphates},
  author={Smith, SH and Wanklyn, BM},
  journal={Journal of Crystal Growth},
  volume={21},
  number={1},
  pages={23--28},
  year={1974},
  publisher={Elsevier}
}

@article{oka2006crystal,
  title={Crystal growth of rare-earth orthovanadate (RVO4) by the floating-zone method},
  author={Oka, Kunihiko and Unoki, Hiromi and Shibata, Hajime and Eisaki, Hiroshi},
  journal={Journal of crystal growth},
  volume={286},
  number={2},
  pages={288--293},
  year={2006},
  publisher={Elsevier}
}

@article{FISCHER199179,
title = {AC method for measuring the magnetocaloric effect and an application in a study on GdVO4},
journal = {Journal of Magnetism and Magnetic Materials},
volume = {94},
number = {1},
pages = {79-84},
year = {1991},
issn = {0304-8853},
doi = {https://doi.org/10.1016/0304-8853(91)90115-Q},
url = {https://www.sciencedirect.com/science/article/pii/030488539190115Q},
author = {B. Fischer and J. Hoffmann and H.G. Kahle and W. Paul}
}

@article{Tokiwa2011,
    author = {Tokiwa, Y. and Gegenwart, P.},
    title = "{High-resolution alternating-field technique to determine the magnetocaloric effect of metals down to very low temperatures}",
    journal = {Review of Scientific Instruments},
    volume = {82},
    number = {1},
    pages = {013905},
    year = {2011},
    month = {01},
    issn = {0034-6748},
    doi = {10.1063/1.3529433},
}

@article{ALIEV2016601,
title = {Magnetocaloric effect in some magnetic materials in alternating magnetic fields up to 22 Hz},
journal = {Journal of Alloys and Compounds},
volume = {676},
pages = {601-605},
year = {2016},
issn = {0925-8388},
doi = {https://doi.org/10.1016/j.jallcom.2016.03.238},
author = {A.M. Aliev and A.B. Batdalov and L.N. Khanov and V.V. Koledov and V.G. Shavrov and I.S. Tereshina and S.V. Taskaev}
}

@article{ALIEV2022169300,
title = {Magnetocaloric effect in manganites in alternating magnetic fields},
journal = {Journal of Magnetism and Magnetic Materials},
volume = {553},
pages = {169300},
year = {2022},
issn = {0304-8853},
doi = {https://doi.org/10.1016/j.jmmm.2022.169300}
}

@article{Warburg1881,
author = {Warburg, E.},
title = {Magnetische Untersuchungen},
journal = {Annalen der Physik},
volume = {249},
number = {5},
pages = {141-164},
doi = {https://doi.org/10.1002/andp.18812490510},
year = {1881}
}

@article{Bruck_2005,
doi = {10.1088/0022-3727/38/23/R01},
url = {https://dx.doi.org/10.1088/0022-3727/38/23/R01},
year = {2005},
month = {nov},
publisher = {},
volume = {38},
number = {23},
pages = {R381},
author = {Brück, Ekkes},
title = {Developments in magnetocaloric refrigeration},
journal = {Journal of Physics D: Applied Physics}
}

@article{Law2018,
author = {Law, Jia Yan and Franco, Victorino and Moreno-Ram{\'{i}}rez, Luis Miguel and Conde, Alejandro and Karpenkov, Dmitriy Y and Radulov, Iliya and Skokov, Konstantin P and Gutfleisch, Oliver},
doi = {10.1038/s41467-018-05111-w},
issn = {2041-1723},
journal = {Nature Communications},
number = {1},
pages = {2680},
title = {{A quantitative criterion for determining the order of magnetic phase transitions using the magnetocaloric effect}},
url = {https://doi.org/10.1038/s41467-018-05111-w},
volume = {9},
year = {2018}
}

@article{Pereira2024,
title = {Complete thermodynamic characterization of second-order phase transition magnetocaloric materials exclusively through magnetometry},
journal = {Journal of Alloys and Compounds},
volume = {976},
pages = {173290},
year = {2024},
issn = {0925-8388},
doi = {https://doi.org/10.1016/j.jallcom.2023.173290},
url = {https://www.sciencedirect.com/science/article/pii/S0925838823045930},
author = {C.S. Pereira and R. Almeida and R. Kiefe and C. Amorim and D.J. Silva and J.S. Amaral and J.H. Belo}
}

@article{Bowden1998,
author = {Bowden, G J},
journal = {Australian Journal of Physics},
number = {2},
pages = {201--236},
title = {{A Review of the Low Temperature Properties of the Rare Earth Vanadates}},
url = {https://doi.org/10.1071/P97066},
volume = {51},
year = {1998}
}

@article{Kaze2006,
  title={Magnetoelastic effects in rare-earth vanadates YbVO4 and HoVO4},
  author={Z. A. Kazeǐ and R. I. Chanieva},
  journal={Journal of Experimental and Theoretical Physics},
  year={2006},
  volume={102},
  pages={266-276},
  url={https://api.semanticscholar.org/CorpusID:122564464}
}

@article{Palacios2018,
  title = {Magnetic structures and magnetocaloric effect in $R{\mathrm{VO}}_{4}(R=\mathrm{Gd}, \mathrm{Nd})$},
  author = {Palacios, E. and Evangelisti, M. and S\'aez-Puche, R. and Dos Santos-Garc\'{\i}a, A. J. and Fern\'andez-Mart\'{\i}nez, F. and Cascales, C. and Castro, M. and Burriel, R. and Fabelo, O. and Rodr\'{\i}guez-Velamaz\'an, J. A.},
  journal = {Phys. Rev. B},
  volume = {97},
  issue = {21},
  pages = {214401},
  numpages = {10},
  year = {2018},
  month = {Jun},
  publisher = {American Physical Society},
  doi = {10.1103/PhysRevB.97.214401},
  url = {https://link.aps.org/doi/10.1103/PhysRevB.97.214401}
}

@article{Khansili2023,
  title = {Calorimetric measurement of nuclear spin-lattice relaxation rate in metals},
  author = {Khansili, A. and Bangura, A. and McDonald, R. D. and Ramshaw, B. J. and Rydh, A. and Shekhter, A.},
  journal = {Phys. Rev. B},
  volume = {107},
  issue = {19},
  pages = {195145},
  numpages = {8},
  year = {2023},
  month = {May},
  publisher = {American Physical Society},
  doi = {10.1103/PhysRevB.107.195145},
  url = {https://link.aps.org/doi/10.1103/PhysRevB.107.195145}
}

@article{Mangum1972,
    author = {Mangum, B. W. and Thornton, D. D.},
    title = "{MAGNETIC PHASE DIAGRAMS OF GdVO4 AND GdAs04}",
    journal = {AIP Conference Proceedings},
    volume = {5},
    number = {1},
    pages = {311-315},
    year = {1972},
    month = {03},
    issn = {0094-243X},
    doi = {10.1063/1.3699447},
    url = {https://doi.org/10.1063/1.3699447}
}

@article{Orbach1961,
 ISSN = {00804630},
 URL = {http://www.jstor.org/stable/2414099},
 author = {R. Orbach},
 journal = {Proceedings of the Royal Society of London. Series A, Mathematical and Physical Sciences},
 number = {1319},
 pages = {458--484},
 publisher = {The Royal Society},
 title = {Spin-Lattice Relaxation in Rare-Earth Salts},
 urldate = {2024-06-27},
 volume = {264},
 year = {1961}
}

@article{Vallipuram2024,
  title = {Role of magnetic ions in the thermal Hall effect of the paramagnetic insulator ${\mathrm{TmVO}}_{4}$},
  author = {Vallipuram, Ashvini and Chen, Lu and Campillo, Emma and Mezidi, Manel and Grissonnanche, Ga\"el and Boulanger, Marie-Eve and Lefran\ifmmode \mbox{\c{c}}\else \c{c}\fi{}ois, \'Etienne and Zic, Mark P. and Li, Yuntian and Fisher, Ian R. and Baglo, Jordan and Taillefer, Louis},
  journal = {Phys. Rev. B},
  volume = {110},
  issue = {4},
  pages = {045144},
  numpages = {7},
  year = {2024},
  month = {Jul},
  publisher = {American Physical Society},
  doi = {10.1103/PhysRevB.110.045144},
  url = {https://link.aps.org/doi/10.1103/PhysRevB.110.045144}
}

@article{tmvo42022,
  title={Field-tuned ferroquadrupolar quantum phase transition in the insulator TmVO4},
  author={Massat, Pierre and Wen, Jiajia and Jiang, Jack M and Hristov, Alexander T and Liu, Yaohua and Smaha, Rebecca W and Feigelson, Robert S and Lee, Young S and Fernandes, Rafael M and Fisher, Ian R},
  journal={Proceedings of the National Academy of Sciences},
  volume={119},
  number={28},
  pages={e2119942119},
  year={2022},
  publisher={National Acad Sciences},
url={https://www.pnas.org/doi/full/10.1073/pnas.2119942119}
}

@article{Santos2007,
    author = {Santos, C. C. and Silva, E. N. and Ayala, A. P. and Guedes, I. and Pizani, P. S. and Loong, C.-K. and Boatner, L. A.},
    title = "{Raman investigations of rare earth orthovanadates}",
    journal = {Journal of Applied Physics},
    volume = {101},
    number = {5},
    pages = {053511},
    year = {2007},
    month = {03},
    issn = {0021-8979},
    doi = {10.1063/1.2437676},
    url = {https://doi.org/10.1063/1.2437676}

}

@article{SANTOS2007_2,
title = {Low-temperature Raman spectra of YbVO4},
journal = {Vibrational Spectroscopy},
volume = {45},
number = {2},
pages = {95-98},
year = {2007},
note = {Raman Spectroscopy Workshop 2006},
issn = {0924-2031},
doi = {https://doi.org/10.1016/j.vibspec.2007.05.002},
url = {https://www.sciencedirect.com/science/article/pii/S0924203107000501},
author = {C.C. Santos and I. Guedes and C.-K. Loong and L.A. Boatner}
}

@article{NIPKO1997,
title = {Crystal field splitting and anomalous thermal expansion in YbVO4},
journal = {Journal of Alloys and Compounds},
volume = {250},
number = {1},
pages = {569-572},
year = {1997},
issn = {0925-8388},
doi = {https://doi.org/10.1016/S0925-8388(96)02565-0},
url = {https://www.sciencedirect.com/science/article/pii/S0925838896025650},
author = {J.C. Nipko and C.-K. Loong and S. Kern and M.M. Abraham and L.A. Boatner}
}

@article{Quilliam2008,
  title = {Evidence of Spin Glass Dynamics in Dilute ${\mathrm{LiHo}}_{x}{\mathbf{Y}}_{1\ensuremath{-}x}{\mathbf{F}}_{4}$},
  author = {Quilliam, J. A. and Meng, S. and Mugford, C. G. A. and Kycia, J. B.},
  journal = {Phys. Rev. Lett.},
  volume = {101},
  issue = {18},
  pages = {187204},
  numpages = {4},
  year = {2008},
  month = {Oct},
  publisher = {American Physical Society},
  doi = {10.1103/PhysRevLett.101.187204},
  url = {https://link.aps.org/doi/10.1103/PhysRevLett.101.187204}
}

@book{Landau,
  author = {Landau,A. and Lifshitz, E. M.},
  year = {1980},
  title = {Statistical Physics, Volume 5, Chapter 110},
  publisher = {Butterworth-Heinemann}
}

@book{Abragambook,
  author = {Abragam, L. D.},
  year = {1961},
    title = {The Principles of Nuclear Magnetism},
  publisher = {Oxford University Press, Oxford}
}

@article{Onsager,
  title = {Reciprocal Relations in Irreversible Processes. I.},
  author = {Onsager, Lars},
  journal = {Phys. Rev.},
  volume = {37},
  issue = {4},
  pages = {405--426},
  numpages = {0},
  year = {1931},
  month = {Feb},
  publisher = {American Physical Society},
  doi = {10.1103/PhysRev.37.405},
  url = {https://link.aps.org/doi/10.1103/PhysRev.37.405}
}

@article{Quilliam2011,
  title = {Dynamics of the magnetic susceptibility deep in the Coulomb phase of the dipolar spin ice material Ho${}_{2}$Ti${}_{2}$O${}_{7}$},
  author = {Quilliam, J. A. and Yaraskavitch, L. R. and Dabkowska, H. A. and Gaulin, B. D. and Kycia, J. B.},
  journal = {Phys. Rev. B},
  volume = {83},
  issue = {9},
  pages = {094424},
  numpages = {10},
  year = {2011},
  month = {Mar},
  publisher = {American Physical Society},
  doi = {10.1103/PhysRevB.83.094424},
  url = {https://link.aps.org/doi/10.1103/PhysRevB.83.094424}
}

@manual{SR860,
  title        = {SR860 Lock-In Amplifier User Manual},
  organization = {Stanford Research Systems},
  year         = {2023},
  note         = {Revision 2.1},
  url          = {https://www.thinksrs.com/products/sr860.html}
}

\end{document}